\documentclass[10pt,journal]{IEEEtran}

\usepackage{cite}
\usepackage{amsmath,amssymb,amsfonts}
\usepackage{subfigure}
\usepackage{booktabs}
\usepackage{xcolor}
\usepackage{graphicx}
\usepackage{multicol}
\usepackage{multirow}
\usepackage{hyperref}
\usepackage{tabularx}
\usepackage[noindentafter]{titlesec}
\usepackage{dsfont}
\usepackage{algorithm}
\usepackage[noend]{algpseudocode}
\usepackage{enumitem}
\usepackage{textcomp}
\usepackage[utf8]{inputenc}
\graphicspath{ {images/} }

\usepackage{mathtools}
\DeclareUnicodeCharacter{2212}{-}
\DeclarePairedDelimiter\norm{\lVert}{\rVert}
\newcommand{\blue}[1]{\textcolor{black}{#1}}

\begin{document}
\bstctlcite{IEEEexample:BSTcontrol}
\urlstyle{tt}
\title{DRAGON: Decentralized Fault Tolerance in Edge Federations}

\author{
        Shreshth~Tuli,
        Giuliano~Casale
    and~Nicholas~R.~Jennings
\IEEEcompsocitemizethanks{
\IEEEcompsocthanksitem S. Tuli, G. Casale and N. R. Jennings are with the Department
of Computing, Imperial College London, United Kingdom.\protect
\IEEEcompsocthanksitem N. R. Jennings is also with Loughborough University, United Kingdom.\protect\\
E-mails: \{s.tuli20, g.casale\}@imperial.ac.uk, n.r.jennings@lboro.ac.uk.\protect
}
\thanks{Manuscript received ---; revised ---.}
\thanks{A preliminary version of the GON framework was presented at the Machine Learning for Systems Workshop at NeurIPS 2021~\cite{tuli2021generative}. The code and datasets are available at \url{https://github.com/imperial-qore/GON}.}}

\markboth{IEEE Transactions on Network and Service Management}%
{Tuli \MakeLowercase{\textit{et al.}}: --- }

\maketitle
\thispagestyle{plain}
\pagestyle{plain}


\begin{abstract}
Edge Federation is a new computing paradigm that seamlessly interconnects the resources of multiple edge service providers.  A key challenge in such systems is the deployment of latency-critical and AI based resource-intensive applications in constrained devices. 
To address this challenge, we propose a novel memory-efficient deep learning based model, namely generative optimization networks (GON). Unlike GANs, GONs use a single network to both discriminate input and generate samples, significantly reducing their memory footprint. Leveraging the low memory footprint of GONs, we propose a decentralized fault-tolerance method called DRAGON that runs simulations (as per a digital modeling twin) to quickly predict and optimize the performance of the edge federation. Extensive experiments with real-world edge computing benchmarks on multiple Raspberry-Pi based federated edge configurations show that DRAGON can outperform the baseline methods in fault-detection and Quality of Service (QoS) metrics. Specifically, the proposed method gives higher F1 scores for fault-detection than the best deep learning (DL) method, while consuming lower memory than the heuristic methods. This allows for improvement in energy consumption, response time and service level agreement violations by up to 74, 63 and 82 percent, respectively. 
\end{abstract}

\begin{IEEEkeywords}
Edge Computing, Decentralized Management, Fault-Tolerance, Federated Computing, Generative Optimization Networks.
\end{IEEEkeywords}

\section{Introduction}\label{sec:introduction}
\noindent
\IEEEPARstart{F}{ederated} computing is a paradigm that brings together the software, infrastructure and platform services from disparate and distributed computing environments. \blue{This prevents resource under-utilization and facilitates the accommodation of sudden spikes in demand~\cite{rochwerger2009reservoir}.} In the past few years, this paradigm has been extended to cloud computing, giving rise to cloud federations~\cite{fdmr}. \blue{However, bringing the computation closer the users, \textit{i.e.}, the edge of the network, has many benefits including lower latency, efficient communication and reduced costs~\cite{gill2019transformative}.} Bolstered by affordable computation, recently proposed federated edge architectures enable large-scale deployment by leveraging multiple standalone edge computing providers~\cite{cao2020edge}. The cost benefits of this seamless interconnect of several edge devices in a federated setup have pushed the industry from relying on cloud-only and hybrid edge-cloud deployments to edge-only ones~\cite{zeydan2016big}. \blue{Due to the large number of devices at the edge, this has led to an explosion in the amount of operational log data being generated. As modern applications shift to resource-intensive AI based workloads, mission critical tasks make it crucial to detect and recover from faults arising from malicious attacks, intrusion events or unexpected device breakdowns by mining this log data~\cite{sharma2019anomaly}.} 

\textbf{Challenges.} Modern edge computing systems now support sophisticated Internet of Things (IoT) and AI based workloads that put excessive load on the limited working memory (up to 8GB) of edge devices~\cite{zhang2019masm, harish2020implementing, hao2018edge}. In such systems, it is often possible to handle the processing limitations by effective preemption and prolonged job execution. However, memory bottlenecks are much harder to solve due to the limited performance of virtual memory in network attached storage settings~\cite{shao2020communication}. This problem is further exacerbated by the latency constraints of cloud backends that often make them unsuitable for offloading IoT jobs~\cite{mukherjee2021amas}. \blue{Without cloud backends, overloaded edge devices make it hard to deploy deep learning models.  All these issues lead to persistent resource contention and faulty behavior adversely affecting system QoS.} 

\begin{table*}[t]
    \centering
    \caption{Comparison of related works with different parameters ($\checkmark$ means that the corresponding feature is present).}
    \label{tab:rw}
    \resizebox{\linewidth}{!}{
\begin{tabular}{@{}lrcccccccc@{}}
\toprule 
Work & Edge & Cloud & Approach & Fault Detection & Fault Remediation & Remediation Type & Scalable & Memory Efficient\tabularnewline
\midrule
Medusa\cite{medusa} &  & \checkmark & Task replication across federations &  & \checkmark & Proactive &  & \checkmark\tabularnewline
PBFM\cite{pbfm} &  & \checkmark & Linear Programming based load balancing &  & \checkmark & Proactive &  & \checkmark\tabularnewline
IoTEF\cite{iotef} & \checkmark & \checkmark & Task virtualization and replication &  & \checkmark & Proactive &  & \checkmark\tabularnewline
TBAFT\cite{tbaft} &  & \checkmark & Overload prediction and task replication & \checkmark & \checkmark & Proactive &  & \checkmark\tabularnewline
AWGG\cite{awgg} &  & \checkmark & Autoencoder based clustering & \checkmark &  & Reactive & \checkmark & \tabularnewline
TopoMAD\cite{topomad} &  & \checkmark & Topology based autoencoder and LSTM & \checkmark &  & Reactive & \checkmark & \tabularnewline
StepGAN\cite{stepgan} &  & \checkmark & Generative Adversarial Network & \checkmark &  & Reactive & \checkmark & \tabularnewline
\textbf{This work} & \checkmark &  & GON and Simulation based preemptive migration & \checkmark & \checkmark & Proactive & \checkmark & \checkmark\tabularnewline
\bottomrule
\end{tabular}}
\end{table*}

\textbf{Existing solutions}. Several fault-tolerance approaches have been proposed for both cloud and edge systems that use heuristics or unsupervised learning strategies~\cite{medusa, tbaft, pbfm, gan2020sage, topomad}. Compared to supervised learning methods that rely on labeled data, unsupervised methods enable fault detection without human annotations or time-consuming log-trace analyses. In this domain, the best performing solutions are contemporary models that leverage memory-intensive deep generative models such as autoencoders (AE), recurrent neural networks and generative adversarial networks (GANs)~\cite{khanna2020intelligent, mad_gan, topomad, girish2021anomaly, chouliaras2021detecting, tuli2021pregan, gan2020sage, stepgan}. State-of-the-art fault-tolerance approaches in the cloud use deep learning due to the ability of such models to capture patterns and trends in large amounts of data. On the other hand, the approaches designed for edge systems are dominated by either heuristic based methods~\cite{grover2018reliable, mudassar2020decentralized} or techniques that use powerful cloud devices as management nodes~\cite{javed2018cefiot, wang2021privacy}. This is particularly due to the memory limitations at the edge. 

The models that have the best detection scores (AEs and GANs) often have a very high memory footprint, 4-6 GB for training and inference in deep learning based methods. 
Due to the large amount of time-series data generated by edge systems~\cite{usad}, the AEs and GANs used in such methods have extremely high parameter size. Some prior work aims at reducing the on-device parameter size of such models using techniques like model compression~\cite{mc}, neural network slimming~\cite{gan_slimming} or model distribution~\cite{li2021model}. However, such approaches usually have accuracy penalties associated with them. This creates a research gap of achieving fault-detection performance comparable with  that of state-of-the-art deep generative models,  within memory footprint constraints.

\textbf{New insights and contributions.} To develop a memory-efficient generative model, we propose a novel model called generative optimization networks (GON). We take motivation from GANs that utilize two neural networks, generator and discriminator, where the former generates new data and the latter acts as a critic to train the former~\cite{goodfellow2014generative, tuli2022carol}. The key insight of our work is that a sufficiently trained discriminator should not only indicate whether an input belongs to a data distribution, but also how to modify the input to make it resemble the target distribution more closely. This can be done using the gradient of the discriminator output with respect to its input~\cite{tuli2021cosco} and executing gradient optimization of the output score. Having only a single network should give significant memory footprint gains, at the cost of higher training times. This is acceptable in many applications scenarios where initial model training is a one-time activity~\cite{goli2020migrating, huang2020clio}. The GON model is agnostic to the specific use case and can be utilized in any application where GANs are used. A preliminary version of the GON model was presented at the Machine Learning for Systems Workshop at NeurIPS 2021~\cite{tuli2021generative}. Compared to the earlier work, we present a full framework namely \underline{D}ecentralized Fault-Tole\underline{ra}nce using \underline{G}enerative \underline{O}ptimization \underline{N}etworks (DRAGON) \blue{that extends GONs using graph neural networks and attention operations with also utilizing a simulator to estimate future system states.}

The decentralized fashion of fault-detection and remediation in DRAGON has a two-fold benefit. Firstly, there is no single point of failure in the system as the detection and remediation steps are run in each edge provider's local edge infrastructure (LEI). Secondly, it allows the distribution of fault-tolerance related management workload across multiple computing devices, facilitating scalability of the model. Specifically, each LEI runs GON in its broker to detect faults within its edge devices. \blue{DRAGON extends the GON model to \textbf{also consider the system topology} and leverages the updated model in a decentralized fashion. It additionally uses a simulator, akin to a cyber-cyber digital-twin, to converge to fault-remediation decisions.}  All brokers altruistically transfer and resume tasks from faulty nodes to a different device for load-balancing, commonly referred to as \textit{preemptive migration}~\cite{engelmann2009proactive}.  The GON model is trained using log traces on DeFog benchmarks~\cite{mcchesney2019defog} and executed on AIoT benchmarks~\cite{luo2018aiot} to test efficacy in unseen settings. DRAGON is the first system that uses a GON in edge brokers to predict faults in edge devices of each LEI and proactively runs load-balancing preemptive migration steps to remediate faults. Extensive experiments on a Raspberry-Pi based federated edge cluster show that DRAGON performs {best} in terms of fault-detection and QoS metrics. To test the generalization of the model, we test on three different federated configurations. Specifically, DRAGON reduces energy consumption, response time and service deadline violations by up to 74\%, 63\% and 82\% compared to state-of-the-art baselines. 

\textbf{Article structure.} Section~\ref{sec:related_work} overviews related work. Section~\ref{sec:motivation} provides a motivating example for the problem. Section~\ref{sec:method} outlines the DRAGON methodology for model training and the optimization of the scheduling decisions. A performance evaluation of the proposed method is presented in Section~\ref{sec:experiments}. Finally, Section~\ref{sec:conclusion} concludes.

\section{Related Work}
\label{sec:related_work}

We now analyze the prior work in fault-tolerant computing. We divide the prior approaches into two categories: heuristic/meta-heuristic methods and deep learning methods. \blue{A comparison is presented in Table~\ref{tab:rw}.}

\textbf{Heuristic and Regression Methods.} Most fault prevention techniques for cloud and edge computing employ some form of heuristics or meta-heuristic approaches. Some techniques like Medusa~\cite{medusa} provide fault-tolerance by creating multiple replicas of each running task across multiple cloud or edge providers allowing resilience against task failures.  Other approaches like Threshold-Based Adaptive Fault Tolerance (TBAFT)~\cite{tbaft} proactively migrate tasks from overloaded devices. The latter uses an ARIMA model to predict resource utilization metrics of host machines and checks them against preset thresholds to execute remediation steps like task migration and replication. \blue{However, in edge computing environments, having node or task level redundancy is inefficient and counterproductive for low-cost deployments~\cite{bagchi2019dependability}. Thus, methods that do not react specifically to faulty scenarios give rise to significant overhead of replication all or most of the incoming tasks.} Another method proposes a preference based fault management technique  (PBFM)~\cite{pbfm} that tries to balance between the QoS improvement and the migration cost using a multi-objective integer linear programming formulation.  Federated Distributed MapReduce (FDMR)~\cite{fdmr} is another MapReduce based framework that leveraged integer linear programming for fault-tolerant distributed task scheduling. \blue{Such methods do not scale for real-time operations, making them unsuitable for mission critical edge applications. This is because common heuristics have been shown to poorly model large-scale or dynamic systems~\cite{tuli2021cosco}.} Another related work for federated edge-cloud computing environments is the IoTEF~\cite{iotef} framework that leverages task virtualization and rerun strategies to avoid service disruption. \blue{All methods in this category (TBAFT, PBFM and FDMR) are memory efficient, remediate fault proactively (by replication or load-balancing) and have lower accuracy than contemporary deep learning based methods as we show in Section~\ref{sec:experiments}.} We use the best performing methods PBFM, IoTEF and TBAFT as baselines in our experiments, as per prior work~\cite{pbfm, iotef, tbaft}.

\textbf{Deep Learning Methods.} Recently, many unsupervised methods have started utilizing deep learning techniques for fault prevention and fault recovery. For instance, the Adaptive Weighted Gath-Geva (AWGG)~\cite{awgg} clustering method is a reconstruction model that detects faults using stacked sparse autoencoders to reduce detection times. Another recent class of methods applies neural networks to reconstruct the last state of the system~\cite{usad, topomad}. The reconstruction error has been used as an indicator of the likelihood of the current state being anomalous~\cite{tuli2022tranad}. For instance, TopoMAD~\cite{topomad} uses a topology-aware neural network that is composed of a Long-Short-Term-Memory (LSTM) and a variational autoencoder (VAE) to detect faults. \blue{However, the reconstruction error is only obtained for the latest state, which limits them to use reactive fault recovery policies.} Other methods use slight variations of LSTM networks with either dropout layers~\cite{girish2021anomaly}, causal Bayesian networks~\cite{gan2020sage} or recurrent autoencoders~\cite{chouliaras2021detecting}. A GAN based approach that uses stepwise training process, StepGAN~\cite{stepgan}, converts the input time-series into matrices and executes convolution operations to capture temporal trends. These methods use various thresholding techniques like Peak Over Threshold (POT)~\cite{siffer2017anomaly} or Kernel Density Estimation (KDE)~\cite{martin1996non}. \blue{However, such techniques are not agnostic to the number of hosts or workloads as they assume a maximum number to the number of the active tasks in the system. Moreover, even with higher accuracy than heuristic based approaches, deep learning models such as deep autoencoders, GANs and recurrent networks are adaptive and accurate, but have high memory footprint that adversely affect system performance.} From this category, we test the above mentioned approaches on the testbed described in Section~\ref{sec:experiments} and use the empirically best techniques as baselines: AWGG, TopoMAD and StepGAN. 

DRAGON is another deep learning based method that aims to resolve the challenge of high resource-utilization of other methods in the same class.

\section{Motivating Example}
\label{sec:motivation}

\begin{figure}
    \centering 
    \includegraphics{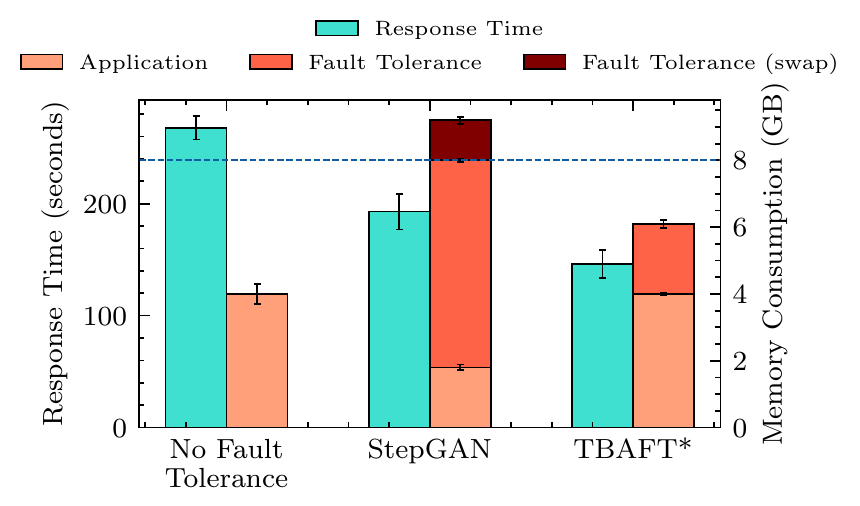}
    \caption{Motivation. Experiments performed on four node edge cluster each with 8GB RAM: (left) No fault tolerance model is run giving high response time and only the application using memory. (middle) StepGAN fault tolerance model that improves response time, but leads to memory contention due to heavy deep neural network. (right) TBAFT heuristic model with precomputed fault labels from StepGAN to obtain memory consumption of TBAFT as a proxy. }
    \label{fig:motivation}
\end{figure}

\begin{figure*}
    \centering
    \includegraphics[width=\linewidth]{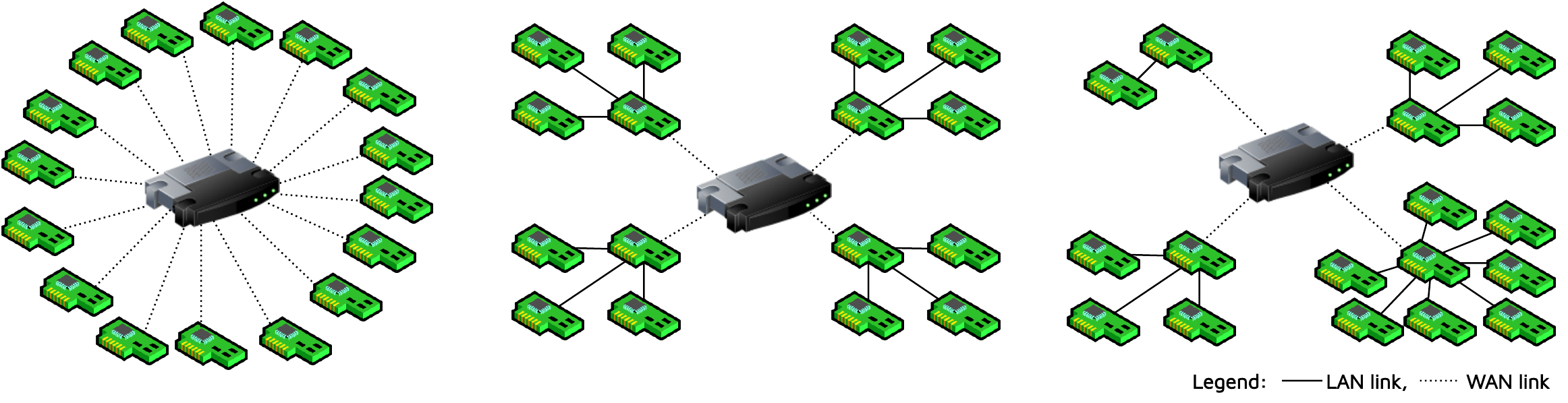}
    \caption{Edge Federation Model with 16 edge nodes. We consider a master-slave architecture of the network where local edge infrastructure (LEI) has a broker for management. All LEIs share resources and interact with each other. The broker of each LEI is connected to the federation using a WAN link to a public-cloud router. (left) 16 homogeneous LEIs with each having only a broker node that also doubles as a worker. (middle) 4 homogeneous LEIs with each broker controlling 4 nodes. (right) 4 heterogeneous LEIs with each broker controlling 2, 4, 4 and 8 workers.}
    \label{fig:system}
\end{figure*}

As demonstrated by prior work~\cite{usad, topomad, awgg}, when increasing the number of edge devices and the amount of operational data, the fault-detection models tend to be more sophisticated. This entails increase in the parameter size of neural networks, subsequently causing higher memory consumption. This load on the limited working memory of edge devices also affects the functioning of the applications running on edge systems. To demonstrate this, we run simple experiments on four Raspberry Pi 4B nodes using the COSCO framework~\cite{tuli2021cosco} and DeFog benchmark applications~\cite{mcchesney2019defog}. Here, the each node has 8GB RAM, acts as its own broker and runs our fault tolerance model in our setup. Fig.~\ref{fig:motivation} shows the results corresponding to three different cases, presenting average response time in blue and memory consumption in red. The first case is when we do not run any fault tolerance model. All the memory of edge nodes is utilized by the running applications in this case. In the second case we run the StepGAN fault-tolerance model, which has a large memory overhead of 7.4GB. Even though it brings down the average response time by 28\%, the model consumes 1.2GB of slow virtual memory (swap space), limiting its optimization capability. In the third case, we run the TBAFT approach, but use precomputed fault labels from the StepGAN model (denoted by TBAFT*). This shows the performance in a case when we have accuracy of the deep learning model with memory-efficiency of a heuristic approach. In this case, as no swap space is utilized, the average response time is much lower, giving 24\% improvement compared to the StepGAN case. This example highlights the potential for an approach that is as memory-efficient as heuristic approaches, but that does not compromise on performance, matching fault detection scores with those of state-of-the-art deep learning techniques.

\section{DRAGON Model}
\label{sec:method}

\subsection{Environment Assumptions and Problem Formulation}
\label{sec:systemmodel}

\textbf{System Model.} We assume a federated edge computing environment with heterogeneous nodes in a master-slave fashion as summarized in Fig.~\ref{fig:system}. Although our methods are generalizable to any federated setup with $H$ hosts, in the rest of the discussion we consider three federated configurations with 16 edge nodes: 
\begin{itemize}[leftmargin=*]
    \item \textit{Configuration 1:} 16 LEI groups, each with a single edge device duplicates as a broker and edge worker. All broker nodes are connected to a Wide Area Network (WAN).
    \item \textit{Configuration 2:} 4 LEI groups, each with its own broker node and 3 worker nodes. All nodes in an LEI are connected with each other in a Local Area Network (LAN).
    \item \textit{Configuration 3:} 4 LEI groups with 2, 4, 4 and 8 nodes. Each LEI has its own broker node.
\end{itemize}
These configurations have been chosen to cover a wide range of heterogeneous federated setups. Configuration 1 and 2 are balanced, whereas configurations 3 is unbalanced in terms of number of nodes in an LEI.

As in traditional federated environments, we assume that all LEI groups are seamlessly connected with each other, allowing information and task sharing across groups. Further, we consider that any execution in this environment lasts for a bounded time. We divide the execution timeline into fixed-sized scheduling intervals, where $I_t$ denotes the $t$-th interval ($t$ ranging from 0 to $T$). We also consider that the edge broker can sample the resource utilization metrics of all hosts within its LEI group at any time (e.g., CPU, RAM, disk/network bandwidth, some additional fault-related metrics including consumption of the swap space, disk buffers, network buffers, disk and network I/O waits~\cite{tanenbaum1997operating}).

\textbf{Fault Model.} We consider different type faults in our system that frequently occur in edge workers while executing data-intensive applications, ranging from breakdowns, malicious attacks and intrusions. We use an existing fault injection module~\cite{ye2018fault} to create different fault types like CPU overload, RAM contention, Disk attack and DDOS attack. The faults manifest in the form of resource over-utilization, for instance a DDOS attack could lead to contention at the network interface. A broker or worker node may become unresponsive due to this resource over-utilization~\cite{vasilakos2020towards}. If a broker breaks down, we allocate the worker with least resource utilization as the broker of that group. As in prior work, edge hosts in the same LEI are connected to the same power supply and unrecoverable faults like outages are ignored~\cite{dftm, javed2018cefiot, bagchi2019dependability}. We prevent over-utilization faults by preemptively migrating tasks from faulty LEIs.

\textbf{Workload Model.} We assume a bag-of-tasks workload model, where a set of independent tasks enter each LEI at the start of each scheduling interval~\cite{tuli2021cosco,tuli2022hunter}. These are generated from the users and are transferred to the edge broker via gateway devices or IoT sensors. 
Each task has an associated Service Level Objective (SLO) deadline. 

\begin{figure*}[t]
    \centering
    \includegraphics[width=0.8\linewidth]{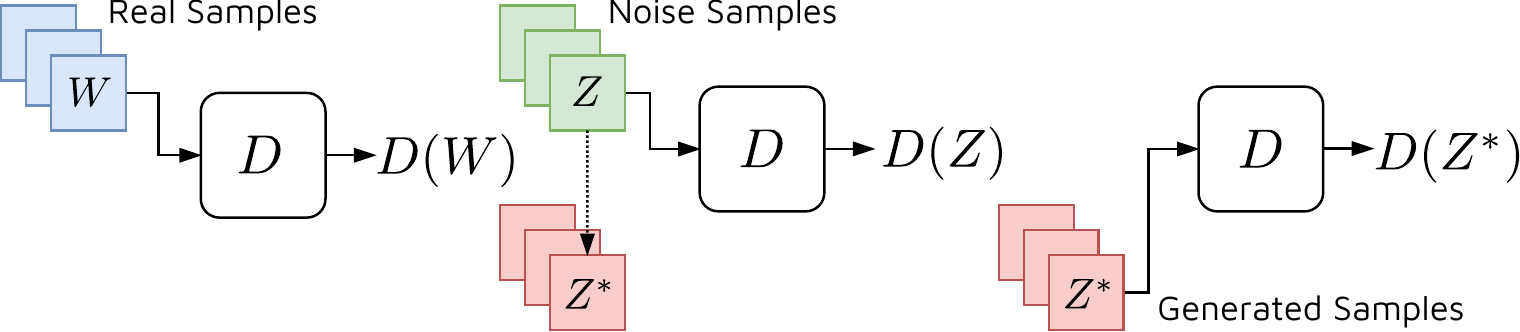}
    \caption{The GON framework. (left) training using real samples from the data. (middle) generating new samples by gradient optimization starting from random noise samples. (right) training using generated samples.}
    \label{fig:framework}
\end{figure*}

\textbf{Formulation.} In this work, we assume that the edge brokers run scheduling algorithms that place each incoming task onto one of the node in the LEI. Apart from this, the brokers also decide tasks that need to be migrated to another LEI as part of fault management decisions. Formally, at the start of the interval $I_t$, the edge broker makes a decision denoted $S_t$, that is a mapping of new tasks in the interval $I_t$ as well as the active tasks from interval $I_{t-1}$ to its edge hosts. The edge broker also makes a decision $M_t$, that migrates active tasks to another LEI. All tasks being migrated to an LEI are considered as incoming tasks in that interval. Unlike new tasks, migratory tasks may resume from a saved state. We also encode the undirected topology graph of the federated edge environment as $G_t$.

At the beginning of interval $I_t$, for an input time-series of LEI system states $\{x_0, \ldots, x_{t}\}$, each broker needs to predict the next system state, \textit{i.e.}, $x_{t+1}$. Here, the system state ($x_t$) consists of values for an arbitrary set of resource utilization metrics for all active tasks and hosts at the start of $I_t$. Instead of directly using the system states, we use a sliding window of length $k$ to capture the temporal contextual trends:
\begin{equation}
\label{eq:window_size}
    W_t = \{x_{t-k+1}, \ldots, x_{t}\}.
\end{equation}
We use replication padding~\cite{liu2018partial} for the first $k$ intervals, where we replicate the first system state to ensure the time-series window is always of the same size. Also, to ensure robust predictions in the DRAGON model, we normalize these time-series windows using min-max scaling.

Using the input window $W_t$, graph topology $G_t$ and an input scheduling decision $S_t$, the model needs to predict whether there is likely to be a fault in the next system state, \textit{i.e.}, $x_{t+1}$. For $S_t$ we assume that there is an underlying scheduler in the system independent from the proposed fault-tolerance solution. In lieu of directly predicting the fault label (denoted as $y_t$), we predict a fault score $f_t$ using a predicted reconstruction of the next window $W_{t+1}$, denoted as $\hat{W}_{t+1}$. The fault score $f_t$ is obtained by calculating the deviation between the true window $W_{t+1}$ and its reconstruction $\hat{W}_{t+1}$. Without loss in generality, whenever unambiguous, we drop the subscripts for the sake of simplicity. Hence, we only refer to the inputs and outputs as $G$, $W$, $S$, $\hat{W}$, and $f$.

\subsection{Generative Optimization Networks}
\label{sec:gon}

We now describe the GON framework for generic data generation and later elucidate its function in fault-detection. In GON, we have a single discriminator model $D(\cdot ; \theta)$ that is a differentiable multilayer perceptron with parameters $\theta$. For any input set of $G, W \text{ and } S$, the output $D(G, W, S; \theta)$ is a scalar. We denote a generated distribution with $p_g$, and with $D(G, W, S; \theta)$ the probability that $W$ belongs to the data $p_{data}$ instead of $p_g$. We train $D$ such that it correctly labels samples from data and $p_g$. A working representation is shown in Fig.~\ref{fig:framework}. The key idea is divided into three parts.

\textbf{Training using real samples.} For any sample $W$ from the data, to make sure the output of $D$ is maximized, we train it to minimize the cross-entropy loss by descending the stochastic gradient
\begin{equation}
\label{eq:l1}
    - \nabla_\theta \log \big(D(G, W, S; \theta) \big).
\end{equation}

\textbf{Generating new samples.} Starting from a random noise sample $Z$, we aim to increase its probability of coming from the data. To do this, we use $D$ as a surrogate model and maximize the objective $\log \big(D(G, Z, S) \big)$. To do this, starting from a random noise sample, a stochastic gradient ascent based solution executes the following until convergence
\begin{equation}
\label{eq:opt}
    Z \gets Z + \gamma \nabla_Z \log \big( D(G, Z, S; \theta) \big).
\end{equation}
Here, $\gamma$ is a step parameter. The optimization loop converges to $Z^*$ that should have a high $D(G, Z^*, S; \theta)$ probability and represents a reconstruction of the next window. The working intuition is that the surrogate surface is highly non-convex and optimization over diverse noise samples would lead to distinct local optima. In principle, every optimum corresponds to a unique element from the data distribution, allowing the GON model to generate a diverse set of samples. Alternatively, we could stop after each step with  probability $D(G, Z, S; \theta)$. A caveat of this approach is that sample generation could be slow as we run optimization in the input space. For some applications this additional overhead might be negligible or acceptable~\cite{goli2020migrating, huang2020clio}.

\textbf{Training using generated samples.} We consider the new samples as self-supervised fake ones~\cite{chen2020self} not belonging to the data and train $D$ using the cross-entropy loss by descending the gradient
\begin{equation}
\label{eq:l2}
     - \nabla_\theta \log \big(1 - D(G, Z^*, S; \theta) \big).
\end{equation}

Informally, as $D$ becomes a better discriminator, minimizing~\eqref{eq:l1} and~\eqref{eq:l2}, the generated samples should also be close to the data. 

\subsection{Optimizing GONs for fault-detection}

The above described GON framework is for generic data generation. However, for the specific case of anomaly/fault detection, we leverage some of its properties to optimize GONs. Consider a GON model that takes inputs the graph topology $G$, time-series window $W$ and scheduling decision $S$ for each LEI. The output of such a GON needs to be a reconstruction of the next window
\[\hat{W} = D(G, W, S; \theta).\]
A bottleneck of the vanilla framework is that the generation of new samples is slow~\cite{tuli2021generative}. To bring down the time to generate new samples, we develop the following optimization.

\textbf{Window Initialization.} Firstly, exploiting the temporal trend of time-series data, instead of starting from a random noise sample $Z$ we can start from the window $W$ and optimize $D(G, Z, S; \theta)$. This is because of the high correlation in the time-series windows of the current and next scheduling intervals in typical edge environments. Thus, we initialize 
\[Z = W.\]

\textbf{Second-Order Optimization.} Secondly, second-order optimization methods have been shown to be much faster for diverse optimization setups~\cite{yao2021adahessian}. So, instead of using a first-order gradient optimization in~\eqref{eq:opt}, we can use higher-order variants. Thus, we could use
\begin{equation}
\label{eq:soopt}
    Z \gets Z + \gamma  [\nabla^2_{Z}\log \big( D(G, Z, S; \theta) \big)]^{-1} \nabla_{Z}\log \big( D(G, Z, S; \theta) \big).
\end{equation}
However, computing the inverse of the second-order gradient (Hessian) is expensive~\cite[\S~6.3]{kochenderfer2019algorithms}. This is alleviated by adapting the Hutchinson's method and using the moving average of the approximate Hessian diagonal to speed up convergence. This is done by using the \texttt{AdaHessian} optimizer~\cite{yao2021adahessian}. We also use warm restarts using cosine annealing~\cite{loshchilov2018decoupled} to speed up convergence.

\begin{figure}
    \centering
    \includegraphics[width=0.65\linewidth]{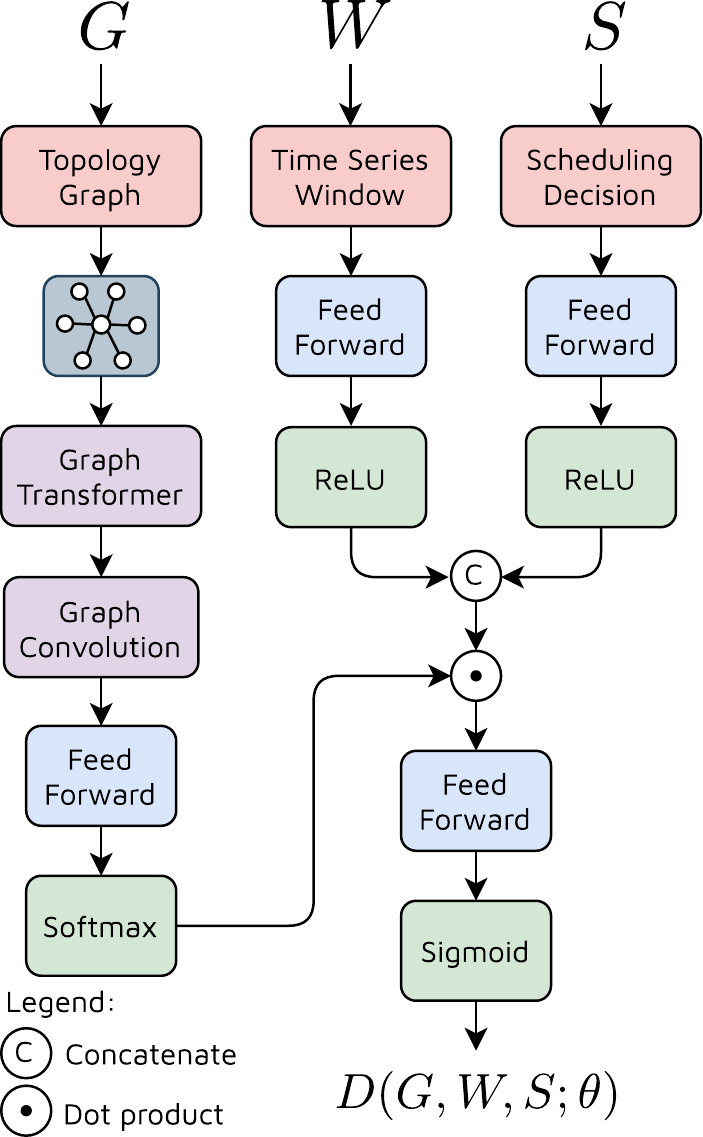}
    \caption{Neural network used in DRAGON. The three inputs to the model are shown in red. Feed-forward, graph operations and activations are shown in blue, purple and green.}
    \label{fig:gon}
\end{figure}

\textbf{GON Network.} It is crucial that the $D$ network is able to capture the temporal trends in the time-series data with the scheduling decision to effectively predict a reconstruction of the next time-series window. The model used in our approach is a composite neural network shown in Fig.~\ref{fig:gon}. Considering we have $p$ tasks running  as a sum of new and active tasks, the scheduling decisions of these are encoded as one-hot vectors of size $m$ (number of hosts in the LEI). We thus get a matrix of scheduling decisions ($S$) of size $[p \times m]$. Also, the $n$ features of $m$ hosts and $p$ tasks in the form of resource utilization metrics, each state window ($W$) is encoded as a $[(m+p) \times n \times k]$ tensor. All operations are performed in a factored fashion, \textit{i.e.}, the neural network operates on $S$ as a batch of $p$ vectors, each having dimension $m$, and on $W$ as a batch of size $m+p$ tensors of size $n\times k$. To encode inputs $W$ and $S$, we use feed-forward layers to bring down the dimension size of the input and $\mathrm{ReLU}$ activation:
\begin{align}
\begin{split}
    E^W &= \mathrm{ReLU}(\mathrm{FeedForward}(W)),\\
    E^S &= \mathrm{ReLU}(\mathrm{FeedForward}(S)).
\end{split}
\end{align}

The topology of the edge federation is represented as a graph with nodes as edge hosts and edges corresponding to edge groups. All edge workers are connected via undirected edges to their respective broker, and all brokers are connected to each other broker. The resource utilization characteristics of each edge host are then used to populate the feature vectors of the nodes in the graph. We denote the feature vector of host $h_i$ as $e_i$. We encode this graph $G$ using a graph transformer network~\cite{yun2019graph} for scalable computation over the input graph. The motivation behind using a graph transformer network is to allow computation across cross-LEI paths that are defined as paths from a worker in one LEI to another worker in a different LEI. This is done by modifying the input graph by graph-transformer operations that generate a transformed graph with cross-LEI paths as edges. As we run preemptive migrations across LEIs, this facilities prediction of effects on QoS and resource utilization metrics in the next interval when migrating task along the cross-LEI paths. Thus, we transform the input graph $G$ as
\begin{equation}
    G_1 = \mathrm{GraphTransformer}(G).
\end{equation}
We then pass the graph through a graph convolution network~\cite{zhang2019graph} to capture the inter host dependencies by convolving feature vectors across the cross-LEI paths in $G$, \textit{i.e.}, the edges in $G_1$. Here, the features for host $h_i$ are aggregated over one-step connected neighbors $n(i)$ in the graph over $r$ convolutions, resulting in an embedding $e^{r}_{i}$ for each host node in the graph. Here, $r$ denotes the number of convolution operations and is typically kept less than the number of LEIs. Specifically, this results in \emph{graph-to-graph} updates as: 
\begin{align}
\begin{split}
    e_i^{0} &= \mathrm{tanh} ( W\ e_{i} + b ),\\
    e^q_i &= \sum_{j \in n(i)} W^q e_{j}^{q-1},
\end{split}
\end{align}
where $W^q$ are the parameters of the graph convolution block in Figure~\ref{fig:gon}. The stacked representation for all hosts is represented as $E^H$. Now, we pass this representation through a feed-forward layer with softmax activation that allows us to use this as attention weights
\begin{equation}
    E^H_1 = \mathrm{Softmax}(\mathrm{FeedForward}(E^H)).
\end{equation}

To generate the final window reconstruction, we use dot-product attention~\cite{chorowski2015attention} as
\begin{equation}
    D(G, W, S; \theta) = \mathrm{Sigmoid}(\mathrm{FeedForward}(E^H_1 \cdot [E^W, E^S])).
\end{equation}
Here, the sigmoid function allows us to bring the output in the range $[0, 1]$, the same as that of the normalized true window.

\begin{algorithm}[!t]
    \begin{algorithmic}[1]
    \Require Dataset $\Lambda  = \{W_0, \ldots, W_T\}$
    \For{ number of training iterations }
    \State Sample minibatch of size $m$ window samples $\{W^{(1)}, \ldots, W^{(m)}\}$ from $\Lambda$.\label{line:sample_data}
    \State Sample minibatch of size $m$ noise samples $\{Z^{(1)}, \ldots, Z^{(m)}\}$ such that $Z^{(i)} = W^{(i)}$.\label{line:sample_noise}
    \State Generate new samples $\{Z^{*(1)}, \ldots, Z^{*(m)}\}$ by running the following till convergence
    \begin{equation*}
        Z \gets Z + \gamma  [\nabla^2_{Z}\log \big( D(G, Z, S; \theta) \big)]^{-1} \cdot \nabla_{Z}\log \big( D(G, Z, S; \theta) \big).%
    \end{equation*}\label{line:optimize}\vspace{-10pt}
    \State Update the discriminator $D(\cdot; \theta)$ by ascending the stochastic gradient.
    \begin{equation*}
        \nabla_\theta \frac{1}{m} \sum_{i=1}^m \big[\log \big(D(G, W^{(i)}, S; \theta) \big) + \log \big(1 - D(G, Z^{*(i)}, S; \theta) \big) \big].
    \end{equation*} \label{line:train}
    \EndFor
    \end{algorithmic}
\caption{Minibatch stochastic gradient based training of generative optimization networks. Input is dataset $\Lambda$ and hyperparameters $m$ and $\gamma$. In experiments we use \texttt{AdaHessian} optimizer for line~\ref{line:optimize} and \texttt{Adam} for line~\ref{line:train}.}
\label{alg:gon}
\end{algorithm}

\section{Fault-Detection and Remediation}

\subsection{Offline Model Training}

A more pragmatic implementation of GON training is summarized in Alg.~\ref{alg:gon}, where we combine the first and last steps in Section~\ref{sec:gon} when training $D$ using minibatches. The training procedure runs offline, \textit{i.e.}, before executing any workloads on the federated edge setup, training the neural network in a GPU equipped machine. The input to the training framework is a dataset of time-series windows $\Lambda = \{W_0, \ldots, W_T\}$. We first randomly sample $m$ window samples from $\Lambda$ ($Z^*$s in line~\ref{line:sample_data}). We also initialize noise samples as the window samples (line~\ref{line:sample_noise}). We then perform \texttt{AdaHessian} optimization on the noise samples to obtain generated windows (line~\ref{line:optimize}). Finally, we train the model using a combined cross-entropy loss of generated ($Z^{*(i)}$) and data samples ($W^{(i)}$), as described in Section~\ref{sec:gon} (line~\ref{line:train}):
\begin{equation}
    L = \log \big(D(G, W^{(i)}, S; \theta) \big) + \log \big(1 - D(G, Z^{*(i)}, S) \big).
\end{equation}

\subsection{Decentralized Fault Tolerance using GON}
The working of the DRAGON framework is summarized in Alg.~\ref{alg:dragon}. The algorithm runs in each broker node of the federated edge setup where the neural network is periodically fine-tuned to adapt to changing environment and workload characteristics. We now elucidate the different steps in detail.

\textbf{Fault Detection.} For unsupervised fault detection using the GON framework, for each input datum from a multivariate time-series, we create a reconstruction using a neural network $D$. Now, for an input window $W_t$, we create a reconstruction of the next window $\hat{W}_{t+1}$, using a trained model $D$ and the \texttt{AdaHessian} based version of~\eqref{eq:soopt} (line~\ref{line:fake}). However, to generate an approximation of the next window, we leverage co-simulation~\cite{tuli2021cosco} that enables a model to quickly run discrete-event simulations alongside and synchronously with the running computer system. Similar to digital-twins, this allows us to continuously capture the system state and get an estimate of the utilization metrics of a future state of the system, further allowing proactive fault remediation. Having the scheduling decision $S_t$ (line~\ref{line:schedulingdecision}) and the current window $W_t$, we denote the discrete-event simulation as (line~\ref{line:sim})
\begin{equation}
    W_{t+1} \gets \mathrm{Sim}(W_t, S_t).
\end{equation}

To generate the fault score we only consider upward spikes of the true window from the reconstructed window. This is due to the nature of our state data, \textit{i.e}, resource utilization metrics leading to faults only when there is a sudden increase in CPU/RAM/disk/network consumption. The $\mathrm{ReLU}$ activation is apt for this as it gives a zero fault score when the true utilization metrics are lower than the predicted ones. Thus, as in line~\ref{line:faultscore}
\begin{equation}
    f = \norm{\mathrm{ReLU}(W_{t+1} - \hat{W}_{t+1})}.
\end{equation}

Once we have the fault scores, we generate fault labels using the Peak Over Threshold (POT) method~\cite{siffer2017anomaly}. POT chooses the threshold automatically and dynamically, above which an fault score is classified as a true class label. Thus, our fault label (line~\ref{line:faultlabel}) is calculated as
\begin{equation}
\label{eq:labeling}
    l = \mathds{1}(f \geq POT(W_t)).
\end{equation}

Comparing the resource utilization metrics of each host separately, the edge broker generates a fault-label for each edge device in its LEI. 

\begin{algorithm}[t]
    \begin{algorithmic}[1]
    \Require
    \Statex Trained GON Model $D(\cdot; \theta)$; 
    \Statex Set of hosts $\Omega$
    \State \textbf{for} $t = 0 \text{ to } T$ \textbf{do}
    \State \hspace{0.8em} \textbf{for each} broker \textbf{do}
    \State \hspace{1.5em} $S_t$ from underlying scheduler \label{line:schedulingdecision}
    \State \hspace{1.5em} $W_{t+1} \gets Sim(W_t, S_t)$ \label{line:sim}
    \State \hspace{1.5em} $\hat{W}_{t+1} \gets$ \texttt{AdaHessian}($D(G_t, W_t, S_t; \theta)$) \label{line:fake}
    \State \hspace{1.5em} $f = \norm{\mathrm{ReLU}(W_{t+1} - \hat{W}_{t+1})}$ \label{line:faultscore}
    \State \hspace{1.5em} $l = \mathds{1}(f \geq POT(W_t))$ \label{line:faultlabel}
    \State \hspace{1.5em} Compile time-series windows of all brokers as $W^M_t$\label{line:compile}%
    \State \hspace{1.5em} Faulty hosts $\omega \subseteq \Omega$ that have $l = 1$ \label{line:faultyhosts}
    \State \hspace{1.5em} Randomly initialize migrations from $\omega$ as $M_t$ \label{line:init}
    \State \hspace{1.5em} $M^*_t \gets \mathrm{SimAnnealing}(M_t, \text{Fitness = \eqref{eq:fitness}})$  \label{line:simulatedannealing}
    \State \hspace{1.5em} \textbf{for each} allocation $(t, h)$ in $M^*_t$ \textbf{do} \label{line:migrate1}
    \State \hspace{2.5em} \textbf{if} $t$ in LEI and $h$ not in LEI \textbf{do} \label{line:migrate2}
    \State \hspace{3em} Migrate $t$ to LEI corresponding to $h$ \label{line:migrate3}
    \State \hspace{1.5em} $L = \log D(G, W_{t+1}, S; \theta)  + \log (1\! -\! D(G, \hat{W}_{t+1}, S) )$\label{line:tune1}%
    \State \hspace{1.5em} Fine-tune $D$ using $L$ and \texttt{Adam} \label{line:tune2}
    \end{algorithmic} 
\caption{The DRAGON fault-tolerance algorithm}
\label{alg:dragon}
\end{algorithm}

\textbf{Co-Simulated Fictitious Play.} Now that we predict the edge hosts that could possible have fault in the next interval, we now need to decide the tasks that should be preemptively migrated from these nodes. To do this, we again use the co-simulation technique. A co-simulator can also give us the QoS score of the next state for resource utilization metrics of all nodes instead of the LEI, denoted by $W^M_t$ (line~\ref{line:compile}), and a preemptive migration decision $M_t$. We denote the co-simulated QoS score and the next window by
\begin{align}
\label{eq:qos}
    QoS_{t+1} &\gets \mathrm{Sim}^{\mathrm{QoS}}(W^M_t, M_t),\\
    W^M_{t+1} &\gets \mathrm{Sim}(W^M_t, M_t).
\end{align}
This score could be a convex combination of multiple metrics like SLO violation rate, energy consumption, response time, etc. However, for this work, we only use response time and energy consumption as these are the most important metrics in edge federations~\cite{tuli2021cosco, fdmr, cao2020edge, shao2020communication}. Thus, 
\begin{equation}
    \mathrm{QoS_{t+1}} = \alpha \cdot ART_{t+1} - \beta \cdot AEC_{t+1}.
\end{equation}
Here, $ART_{t+1}$ is the average response time for all tasks leaving the system in interval $I_{t+1}$ normalized by the maximum response time until the current interval. $AEC_{t+1}$ is the average energy consumption of the infrastructure (which includes all edge hosts) in interval $I_{t+1}$ normalized by the maximum power of the hosts.\footnote{Both AEC and ART are unit-less metrics and belong in the range $[0, 1]$. See~\cite[\S~5.1]{tuli2021cosco} for more details.} Also, $\alpha$ and $\beta$ are hyperparameters that can be tuned by the user based on the application requirements. Also, using $W^M_{t+1}$ we detect the number of anomalies in $I_{t+1}$. We denote this by $N_{t+1}$ and use the normalized value ${N_{t+1}}/{H}$, where $H$ is the total number of hosts in the federated environment (in our case $H = 16$). To keep into account the costly migration overhead, we also include the migration overhead score $O_t$ that is the migration time of all tasks in $M_t$ normalized by the maximum migration time until the current interval. We define a fitness score as
\begin{equation}
\label{eq:fitness}
    \text{Fitness}_{t+1}(W^M_t, M_t) = QoS_{t+1} + \delta_1 \cdot \frac{N_{t+1}}{H} + \delta_2 \cdot O_t,
\end{equation}
where $\delta_1$ and $\delta_2$ are hyperparamaters set based on user requirements. The motivation behind this fitness score is that when we optimize a preemptive migration decision, we need to make sure that we consider the immediate reward in terms of the QoS score, but also consider the migration overheads and number of faults in the next interval. This entails the minimization of the fitness score given in~\eqref{eq:fitness}.

Thus, at the start of every interval, each broker node gets the time-series window of other LEIs, combines them to form $W^M_t$ and runs a simulated execution to get the next window and QoS scores. This kind of distributed execution in a multi-agent setup to estimate objective function of a global environment is commonly referred to as ``fictitious play''~\cite{fictitiousplay}. 

\textbf{Preemptive Migration Decision.} The co-simulated fictitious-play described earlier is a very fast and lightweight operation compared to an actual execution of a preemptive migration decision. In fact, running a simulation for $W^M_t$ and $M_T$ is so fast that testing for all possible $M_t$ takes $\sim$0.5 seconds second for 16 nodes. This increases linearly, by around $5\%$ for each new node in the system. Still, for scalability, we run simulated annealing on $M_t$ to optimize the fitness score (line~\ref{line:simulatedannealing}). A neighbor of a decision $M_t$ is defined by either migrating a task not in $M_t$ to another LEI or migrating a task in $M_t$ to a different LEI than in $M_t$. Thus, optimizing fitness score from a preemptive migration decision, which randomly selects tasks to be migrated from faulty hosts detected by~\eqref{eq:labeling} (line~\ref{line:init}), converges to the decision $M^*_t$. This is the output of the DRAGON model. Note that each broker generates such a decision and migrates the set of tasks in $M^*_t$ that are active in its LEI to the target LEI group (lines~\ref{line:migrate1}-\ref{line:migrate3}). All brokers altruistically accept incoming tasks from other LEIs and consider them new tasks in their group. The fault labels are also combined using logical or, for each interval from $t$ to $t+N-1$ to generate the fault label for DRAGON.

\begin{algorithm}[t]
    \begin{algorithmic}[1]
    \color{black}
    \Require
    \Statex Trained GON Model $D(\cdot; \theta)$; 
    \Statex Set of hosts $\Omega$; Look-ahead intervals $N$;
    \State \textbf{for} $t = 0 \text{ to } T$ \textbf{do}
    \State \hspace{0.8em} \textbf{for each} broker \textbf{do}
    \State \hspace{1.5em} $S_t$ from underlying scheduler \label{line:schedulingdecision}
    \State \hspace{1.5em} $W_{t+1} \gets Sim(W_t, S_t)$ \label{line:sim}
    \State \hspace{1.5em} $\hat{W}_{t+1} \gets$ \texttt{AdaHessian}($D(G_t, W_t, S_t; \theta)$) \label{line:fake}
    \State \hspace{1.5em} $f = \norm{\mathrm{ReLU}(W_{t+1} - \hat{W}_{t+1})}$ \label{line:faultscore}
    \State \hspace{1.5em} $l = \mathds{1}(f \geq POT(W_t))$ \label{line:faultlabel}
    \State \hspace{1.5em} Compile time-series windows of all brokers as $W^M_t$\label{line:compile}%
    \State \hspace{1.5em} Faulty hosts $\omega \subseteq \Omega$ that have $l = 1$ \label{line:faultyhosts}
    \State \hspace{1.5em} Initialize empty migration decision as $M^+_t$
    \State \hspace{1.5em} \textbf{for} $k = 0$ to $N$ \textbf{do}
    \State \hspace{2.5em} Randomly initialize migrations from $\omega$ as $M_{t+k}$ \label{line:init}
    \State \hspace{2.5em} $M^*_{t+k}\! \gets\! \mathrm{SimAnnealing}(M_{t+k}, \text{Fitn.\! =\! \eqref{eq:fitness}})$  \label{line:simulatedannealing}%
    \State \hspace{2.5em} $QoS_{t+k+1} \gets \mathrm{Sim}^{\mathrm{QoS}}(W^M_{t+k}, M_{t+k})$ \label{line:updateqos}
    \State \hspace{2.5em} $W^M_{t+k+1} \gets \mathrm{Sim}(W^M_{t+k}, M_{t+k})$ \label{line:updatew}
    \State \hspace{2.5em} \textbf{for each} allocation $(t, h)$ in $M^*_{t+k}$ \textbf{do} \label{line:forstart}
    \State \hspace{4em} \textbf{if} $\exists\ h^+ | (t,h^+) \in M^+_t$ \textbf{do}
    \State \hspace{5em} Replace $(t, h^+)$ with $(t,h)$
    \State \hspace{4em} \textbf{else do}
    \State \hspace{5em} Add $(t,h)$ to $M^+_t$ \label{line:forend}
    \State \hspace{1.5em} \textbf{for each} allocation $(t, h)$ in $M^*_t$ \textbf{do} \label{line:migrate1}
    \State \hspace{2.5em} \textbf{if} $t$ in LEI and $h$ not in LEI \textbf{do} \label{line:migrate2}
    \State \hspace{3em} Migrate $t$ to LEI corresponding to $h$ \label{line:migrate3}
    \State \hspace{1.5em} $L = \log D(G, W_{t+1}, S; \theta)  + \log (1\! -\! D(G, \hat{W}_{t+1}, S) )$\label{line:tune1}%
    \State \hspace{1.5em} Fine-tune $D$ using $L$ and \texttt{Adam} \label{line:tune2}
    \end{algorithmic} 
\caption{The DRAGON+ fault-tolerance algorithm}
\label{alg:dragonplus}
\end{algorithm}

\textbf{Extending to DRAGON+.} As DRAGON just optimizes the fitness score for the next interval, it is myopic in its optimization capacity. To tackle this, we run multiple steps of DRAGON for a long-term optimization strategy and call this DRAGON+ \blue{(summarized in Alg.~\ref{alg:dragonplus}). Assume, for example, that in the first iteration DRAGON generated a decision $M^*_t$. Now, having $W^M_{t+1}$, we can generate $S_{t+1}$, and subsequently $QoS_{t+1}$ from the underlying scheduler and run DRAGON again to generate $M^*_{t+1}$. We continue this for $N$ intervals and combine $\{M^*_t, \ldots, M^*_{t+N-1}\}$.} To do this, if task $t$ is allocated to $h_i$ in $M^*_p$ and $h_j$ in $M^*_q$, then we select $h_{\max(p, q)}$. This is done because a migration to $h_i$ then later to $h_j$ has higher overhead that direct migration to $h_j$. \blue{This is achieved by maintaining a migration decision $M^+_t$ and updating the decision as we iterate from $t$ to $t+N$ (see lines~\ref{line:forstart}-\ref{line:forend}).} 

\blue{As the DRAGON+ approach performs optimization using a long-term QoS estimation, it can make more informed decisions. Further, DRAGON+ avoid unnecessary task migrations by looking at future migrations by DRAGON. However, the level of performance uplift depends on the volatility of the workloads and the look-ahead interval count $N$ (see Section~\ref{sec:impl}). Note that DRAGON is a special case of DRAGON+ where $N = 1$. Thus, for $N$ higher than 1, the decision time and subsequently the scheduling time would increase, possibly affecting the overall performance of the system. A more empirical comparison is shown in Section~\ref{fig:sens_lr}. Overall, for computationally powerful brokers DRAGON+ is an appropriate choice whereas DRAGON is suitable in case of resource constrained brokers.}

\textbf{Model Fine-tuning.} Now that we have a true window $W_{t+1}$ and a generated window $\hat{W}_t$, we can fine-tune the model with this new datapoint using the loss
\begin{equation}
    L = \log \big(D(G, W_{t+1}, S; \theta) \big) + \log \big(1 - D(G, \hat{W}_{t+1}, S) \big).
\end{equation}
This allows the model to dynamically adapt in non-stationary environments (lines~\ref{line:tune1}-\ref{line:tune2}).

To summarize, the novel contribution of this work is two-fold. First, we develop a decentralized fault-detection model based on the GON framework that uses significantly lower number of parameters than previous generative models to detect faults in each LEI independently. Second, we develop a synchronization strategy using co-simulated fictitious play to jointly decide of preemptive migrations that aim at avoiding faults arising in the system. Thus, we run migrations proactively, \textit{i.e.}, predict faults using GONs and simulations, to avoid faulty states. For crash-like states, we restart tasks. 

\section{Experiments}
\label{sec:experiments}

We compare the DRAGON and DRAGON+ methods against three heuristic baselines: PBFM~\cite{pbfm}, IoTEF~\cite{iotef}, TBAFT~\cite{tbaft}, and three DL baselines: AWGG~\cite{awgg}, TopoMAD~\cite{topomad} and StepGAN~\cite{stepgan} (more details in Section~\ref{sec:related_work}). As TopoMAD and AWGG are only fault-detection methods, we supplement them with the multi-objective integer linear programming based preemptive migration scheme for fault-recovery from another baseline, PBFM. We use hyperparameters of the baseline models as presented in their respective papers. We train all deep learning models using the PyTorch-1.8.0~\cite{paszke2019pytorch} library. 

\subsection{Evaluation Setup}
Our evaluation setup consists of 16 Raspberry Pi 4B nodes, 8 with 4GB RAM and another 8 with 8GB RAM each. This allows the setup to have heterogeneous nodes with different memory capacities. We consider the three federated configurations presented in Section~\ref{sec:systemmodel}. In configuration 1, we have 8 brokers with 4GB and other 8 with 8GB RAM. In configuration 2, one LEI has only 4GB nodes, second LEI has only 8GB nodes, the other two have two 4GB and two 8GB nodes with one with 4GB node as broker and other with 8GB node as its broker. In configuration 3, the LEI with 8 nodes has all 8GB nodes. These configurations and distribution of nodes was kept to maximize heterogeneity and consider diverse federated setups. All WAN and LAN links were 1 Gbps. To emulate each LEI being in geographically distant locations, we use the \texttt{NetLimiter} tool to tweak the communication latency among brokers node using the model described in~\cite{gilly2020modelling}.

To generate the tasks in our system, we use the \textit{AIoTBench} applications~\cite{luo2018aiot}. AIoTBench is a AI based edge computing benchmark suite that consists of various real-world computer vision application instances. The seven specific application types correspond to the neural networks they utilize. These include three typical heavy-weight networks: ResNet18, ResNet34, ResNext32x4d, as well as four light-weight networks: SqueezeNet, GoogleNet, MobileNetV2, MnasNet.\footnote{AIoTBench: \url{https://www.benchcouncil.org/aibench/aiotbench/index.html}. Accessed: 5 October 2021.} This benchmark has been used in our experiments due to its volatile utilization characteristics and heterogeneous resource requirements. The benchmark consists of 50,000 images from the COCO dataset as workloads~\cite{coco}. To evaluate the proposed method in a controlled environment, we abstract out the users and IoT layers described in Section~\ref{sec:systemmodel} and use a discrete probability distribution to realize tasks as container instances. Thus, at the start of each scheduling interval, we create new tasks from a Poisson distribution with rate $\lambda = 1.2$, sampled uniformly from the seven applications. The Poisson distribution is a natural choice for a bag-of-tasks workload model, common in edge environments~\cite{mao2016dynamic, basu2019learn}. Our tasks are executed using Docker containers. We run all experiments for 100 scheduling intervals, with each interval being 300 seconds long, giving a total experiment time of 8 hours 20 minutes. We average over five runs and use diverse workload types to ensure statistical significance of our experiments.

\subsection{Evaluation Metrics}

\textbf{Fault Detection.} To evaluate fault-detection efficacy, we utilize commonly used metrics including accuracy, precision, recall, area under the receiver operating characteristic curve (AUROC) and F1 score. We also consider the ratio of F1 score with the memory consumption of the model (F1/GB) to get an estimate of the detection performance normalized by the memory footprint.

\textbf{Fault Diagnosis.} For fault-diagnosis (\textit{i.e.}, the prediction of the specific hosts that have faults), we use the prediction of fault scores and labels for each host independently. We use two popular metrics for comparison: (1) HitRate@100\% is the measure of how many ground truth dimensions have been included in the top candidates predicted by the model~\cite{omnianomaly}, (2) Normalized Discounted Cumulative Gain NDCG@100\%)~\cite{jarvelin2002cumulated}. 

\textbf{QoS.} For QoS, we measure the energy consumption of the federated setup, average response time and SLO violation rate of completed tasks, Jain's fairness index. We consider the relative definition of SLO (as in~\cite{tuli2021cosco}) where the deadline is the 90$^{th}$ percentile response time for the same application on the state-of-the-art method StepGAN that has the highest F1 scores among the baselines. We also consider the number of task migrations, average time per migration. We also compare overheads in terms of scheduling time and memory consumption of all models.

\begin{table}[t]
    \centering
    \caption{Dataset Statistics}
    \begin{tabular}{@{}llrrrr@{}}
    \toprule
    Dataset &  & Train & Test & Dimensions & Anomalies (\%)\tabularnewline
    \midrule
    FTSAD-1 &  & 600 & 5000 & 1 & 12.88\tabularnewline
    FTSAD-25 &  & 574 & 1700 & 25 & 32.23\tabularnewline
    FTSAD-55 &  & 2158 & 2264 & 55 & 13.69\tabularnewline
    \bottomrule
    \end{tabular}
    \label{tab:datasets}
\end{table}

\subsection{Implementation and Training Details}
\label{sec:impl}

\textbf{Extending COSCO.} To implement and evaluate the proposed methods, we need a framework that we can use to deploy containerized workloads on a federated edge computing environment. One such framework is COSCO~\cite{tuli2021cosco}. It enables the development and deployment of integrated edge-cloud environments with structured communication and platform independent execution of applications. The resource management and task initiation is undertaken on edge nodes in the broker layer. The framework uses HTTP RESTful APIs for communication and seamlessly integrates a \texttt{Flask} based web-environment to deploy and manage containers in a distributed setup~\cite{grinberg2018flask}.

It uses the Checkpoint/Restore In Userspace (CRIU)~\cite{venkatesh2019fast} tool for container migration. All sharing of resource utilization characteristics across brokers uses the \texttt{rsync} utility. We extend the \texttt{Framework} class to allow decentralized decision making. For synchronization of outputs and execution of workloads, we utilize the HTTP Notification API.

\textbf{Fog Time Series Anomaly Detection (FTSAD) datasets.} The FTSAD is a suite of datasets collected on a Raspberry Pi 4B compute cluster~\cite{tuli2021generative}. The cluster consists of five 4GB versions and five 8GB versions. We ran the DeFog benchmark applications~\cite{mcchesney2019defog}, specifically the Yolo, PocketSphinx and Aeneas workloads. We use a random scheduler and execute these tasks as Docker containers. We create faults of the type: CPU overload, RAM contention, Disk attack and DDOS attack~\cite{ye2018fault}. In a CPU overload attack, a simple CPU hogging application is executed that create contention of the compute resources. In RAM contention attack, a program is run that performs continuous memory read/write operations. In disk attack, we run the IOZone benchmark that consumes a large portion of the disk bandwidth. In DDOS attack, we perform several invalid HTTP server connection requests causing network bandwidth contention. We generate faults using a Poisson distribution with rate $\lambda_f = 5$, sampled uniformly at random from the attack set. More details are given in~\cite{ye2018fault}.

For the 10-node cluster, we collect traces of resource utilizations including CPU, RAM, disk and network. Each datapoint is collected at the interval of 10 seconds. We truncate the dataset to have only 1, 25 or 55 dimensions and call these FTSAD-1/25/55 datasets. For FTSAD-1, we use the CPU utilization of one of the cluster nodes. For FTSAD-25, we use the CPU and network read bandwidth utilization traces for all 10 nodes and RAM utilization for the five 4GB nodes (as 8GB nodes rarely have memory contentions). For FTSAD-55, we use the CPU utilization, disk read, disk write, network read, network write bandwidths for all nodes and RAM utilization of the five 4GB nodes. The overall traces are collected for 737 minutes, giving 4422 datapoints. For the FTSAD-55 dataset, we split them at the 2158-\textit{th} interval at which we started saving the anomaly labels as well, giving a test set size of 2264 datapoints. The FTSAD-1/25 datasets were manually curated to have low anomaly frequency in FTSAD-1 and high frequency in FTSAD-25. This allowed us to create a collection of datasets with diverse fault characteristics and dataset sizes.  We summarize the characteristics of the three datasets in Table~\ref{tab:datasets}

\begin{figure}
    \centering
    \includegraphics[width=\linewidth]{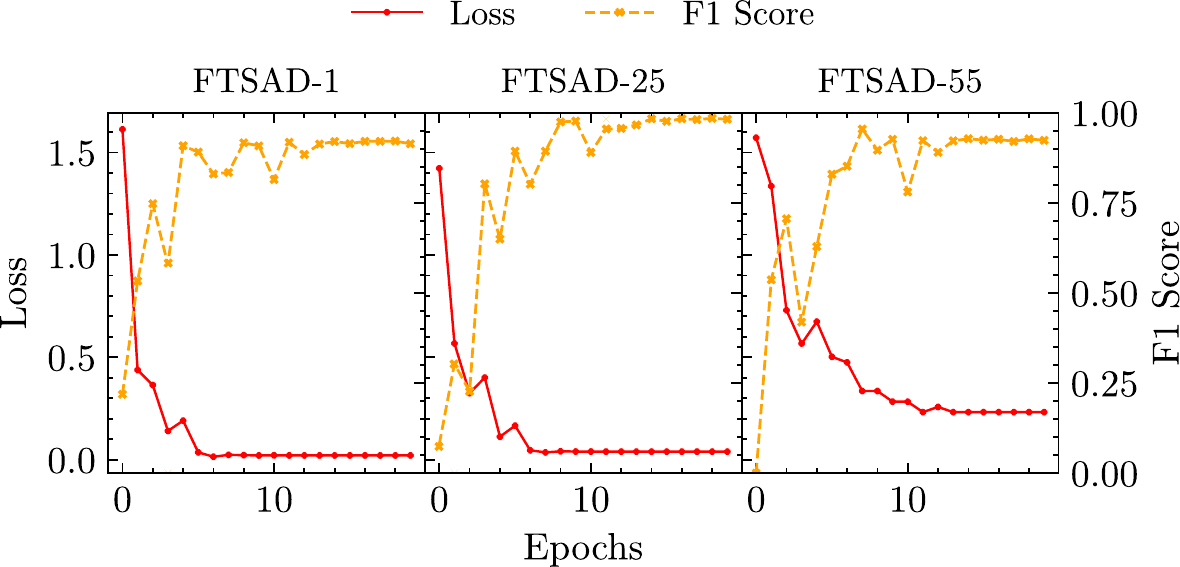}
    \caption{Training the GON model.}
    \label{fig:convergence}
\end{figure}

\textbf{Model Training.} We train the GON network using Alg.~\ref{alg:gon} on the three FTSAD datasets. All model training is performed on a separate server with this configuration: Intel i7 10700k CPU, 16GB RAM, Nvidia RTX 3060 and Windows 11 OS. As the model is agnostic to the dimension size of the multivariate time series, our model can be trained for any of these datasets without changing the network size. For our experiments, we use the model trained on FTSAD-55 as it covers the highest number of parameters. The time-series windows and the scheduling decisions were taken directly from the datasets, whereas the graph topology was taken from the three configurations. For each configuration, we train a separate model. For training we randomly split the dataset into 80\% training and 20\% testing data. We use learning rate of $10^{-4}$ with weight decay of $10^{-5}$. We use \texttt{Adam} optimizer for optimizing the loss function. The training curves showing loss and F1 scores on the test set are shown in Fig.~\ref{fig:convergence}. We use early stopping criterion for convergence. 

\begin{figure}[t]
    \centering
    \includegraphics[width=\linewidth]{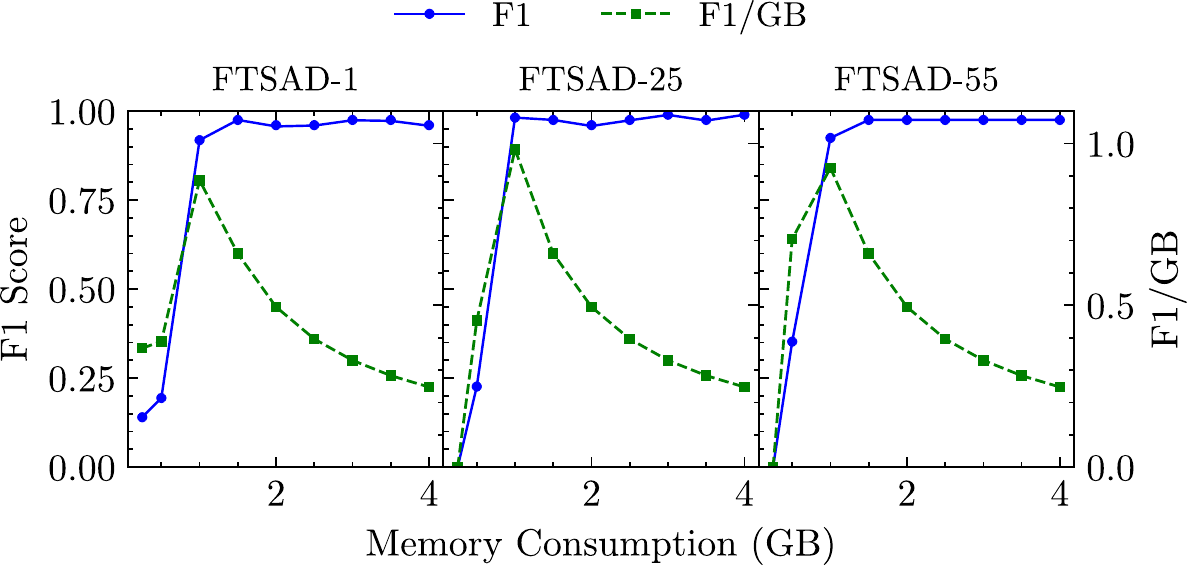}
    \caption{F1 and F1/GB of DRAGON with memory consumption on the FTSAD-1/25/55 datasets.}
    \label{fig:hyper}
\end{figure}
\begin{figure}[t]
    \centering
    \includegraphics[width=\linewidth]{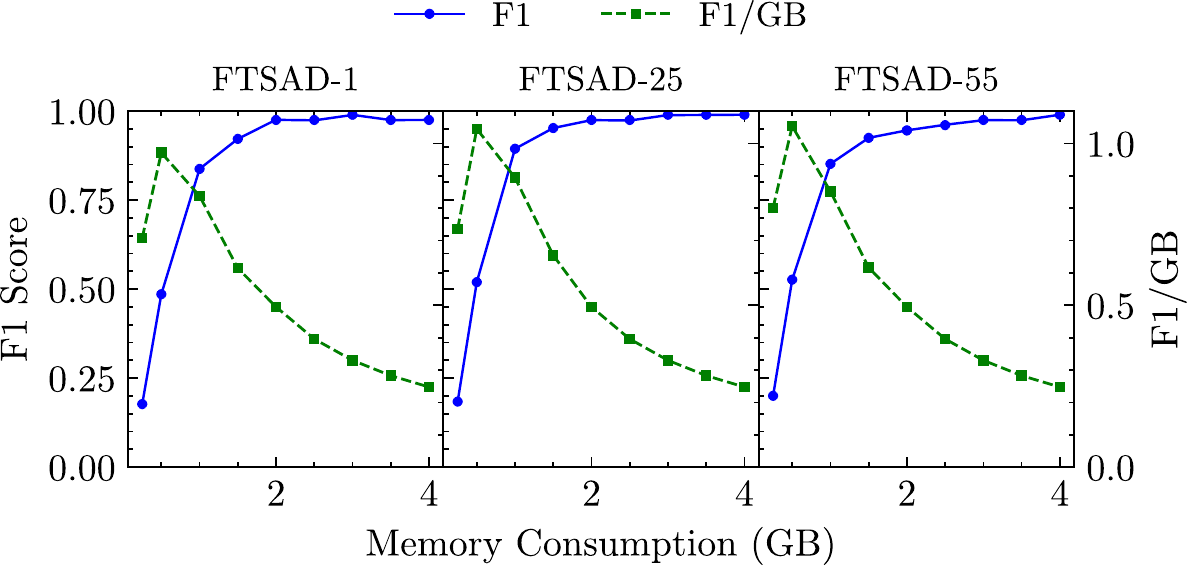}
    \caption{F1 and F1/GB of DRAGON+ with memory consumption on the FTSAD-1/25/55 datasets.}
    \label{fig:hyperplus}
\end{figure}

\textbf{Hyperparameter Selection.} All baseline models have been developed by us for the COSCO framework.  For fault-labeling, we use the following POT parameters, $\mathrm{coefficient} = 10^{-4}$ for all data sets, low quantile is 0.07. These are selected as per~\cite{omnianomaly}. No baseline method provides the flexibility of changing the memory footprint of the trained model. In heuristic models, we use the same parameters as mentioned in their respective papers. However, for AWGG, TopoMAD and StepGAN we can also control the number of parameters by employing model compression or GAN slimming~\cite{mc, gan_slimming}. For GON, we only change the number of layers in the feed-forward networks in Fig.~\ref{fig:gon} (keeping a fixed layer size of 128 nodes). We choose the model memory footprint based on grid search so to maximize F1/GB. The motivation behind this metric is as follows. Increasing memory footprint leads to a monotonic increase in the anomaly detection performance (see Fig.~\ref{fig:hyper} for DRAGON and Fig.~\ref{fig:hyperplus} for DRAGON+). However, when running online fault detection on edge devices, several other applications may be running alongside. Having the maximum possible F1 score would impact negatively the performance of the running applications due to memory contentions. However, having a low memory footprint model but with poor F1 score would be ineffective in detecting anomalies for apt remediation. Thus, a careful balance needs to be maintained to prevent memory contentions and provide high detection. This can be seen by Fig.~\ref{fig:hyper} that shows the F1 and F1/GB scores of DRAGON model on the FTSAD datasets. 

\blue{Figs.~\ref{fig:hyper} and~\ref{fig:hyperplus} show that the accuracy improvement in terms of the F1 score is sharper for DRAGON compared to DRAGON+. This is due to the fact that DRAGON performs fault detection only for the state of the system in the next interval, whereas DRAGON+ considers $N$ intervals. Considering the next step prediction being a simpler problem to model, the neural network in DRAGON can quickly fit the training data without having a large parameter set. DRAGON+ involves multi-step lookahead where the detection performance does not rise as sharply as the memory footprint, and hence the F1/GB peak is seen at 0.5GB for DRAGON+. On the other hand, for DRAGON, model of size $\sim$1GB has the maximum F1/GB score and hence have been used in our experiments. To intuitively understand why the F1/GB peak of DRAGON+ is for a model with a lower memory footprint, we compare the input data. DRAGON gets a single state window $W_t$ to predict faults, whereas DRAGON+ gets $N$ such windows. Having access to more data enables us to inject the system state information using a simulation into DRAGON+, requiring less parameters to fit the model.} Similarly, we determine the compression in AWGG, TopoMAD and the slimming factor in StepGAN.

For next window generation we use \texttt{AdaHessian} optimizer with \texttt{CosineAnnealing} learning rate decay with warm restarts every 10 epochs. We use learning rate ($\gamma$ in Alg.~\ref{alg:gon}) of $10^{-3}$ and window size ($K=10$ in~\eqref{eq:window_size} for DRAGON and $K=5$ for DRAGON+). These values were selected using grid-search to optimize F1 score (more details in Section~\ref{sec:sensitivity}). Other parameters chosen using grid search include: minibatch size $m$ as 10 to optimize F1 score, $\alpha = \beta = 0.5$ in~\eqref{eq:qos} as per prior work~\cite{basu2019learn, tuli2021cosco}, $\delta_1 = \delta_2 = 0.5$ in~\eqref{eq:fitness} and number of iterations $N = 5$ in DRAGON+ to minimize SLO violation rates.

\begin{table*}
\centering
\caption{Comparison of detection and diagnosis scores with standard deviations. Results reported on the test set with 80:20 train-test split. }
\label{tab:detection}
\resizebox{\linewidth}{!}{
\begin{tabular}{@{}lcccccccc@{}}
\toprule 
\multirow{2}{*}{Method} & \multicolumn{6}{c}{Detection} & \multicolumn{2}{c}{Diagnosis}\tabularnewline
\cmidrule{2-9} 
 & Accuracy & Precision & Recall & AUROC & F1 & F1/GB & HR@100 & NDCG@100\tabularnewline
\midrule 
\multicolumn{9}{c}{FTSAD-1}\tabularnewline
\midrule 
PBFM & 0.8526\scriptsize{$\pm$0.0039} & 0.7362\scriptsize{$\pm$0.0015} & 0.8412\scriptsize{$\pm$0.0075} & 0.8713\scriptsize{$\pm$0.0070} & 0.7852\scriptsize{$\pm$0.0031} & 0.7325\scriptsize{$\pm$0.0062} & 0.5461\scriptsize{$\pm$0.0048} & 0.4716\scriptsize{$\pm$0.0031}\tabularnewline
IoTEF & 0.8713\scriptsize{$\pm$0.0014} & 0.7814\scriptsize{$\pm$0.0049} & 0.8721\scriptsize{$\pm$0.0032} & 0.8827\scriptsize{$\pm$0.0048} & 0.8243\scriptsize{$\pm$0.0010} & 0.7418\scriptsize{$\pm$0.0001} & 0.4958\scriptsize{$\pm$0.0049} & 0.5125\scriptsize{$\pm$0.0019}\tabularnewline
TBAFT & 0.8967\scriptsize{$\pm$0.0088} & 0.8012\scriptsize{$\pm$0.0068} & 0.8516\scriptsize{$\pm$0.0037} & 0.8561\scriptsize{$\pm$0.0054} & 0.8257\scriptsize{$\pm$0.0000} & 0.7737\scriptsize{$\pm$0.0055} & 0.5112\scriptsize{$\pm$0.0022} & 0.6019\scriptsize{$\pm$0.0042}\tabularnewline
AWGG & 0.7981\scriptsize{$\pm$0.0039} & 0.7613\scriptsize{$\pm$0.0039} & 0.8913\scriptsize{$\pm$0.0055} & 0.8824\scriptsize{$\pm$0.0056} & 0.8212\scriptsize{$\pm$0.0024} & 0.4599\scriptsize{$\pm$0.0006} & 0.5827\scriptsize{$\pm$0.0044} & 0.6110\scriptsize{$\pm$0.0010}\tabularnewline
TopoMAD & 0.8924\scriptsize{$\pm$0.0083} & 0.8385\scriptsize{$\pm$0.0072} & 0.7123\scriptsize{$\pm$0.0055} & 0.9012\scriptsize{$\pm$0.0033} & 0.7703\scriptsize{$\pm$0.0036} & 0.2095\scriptsize{$\pm$0.0020} & 0.5918\scriptsize{$\pm$0.0028} & 0.6512\scriptsize{$\pm$0.0044}\tabularnewline
StepGAN & 0.8987\scriptsize{$\pm$0.0058} & 0.8622\scriptsize{$\pm$0.0045} & 0.8345\scriptsize{$\pm$0.0011} & 0.9066\scriptsize{$\pm$0.0062} & 0.8481\scriptsize{$\pm$0.0052} & 0.3420\scriptsize{$\pm$0.0017} & 0.5946\scriptsize{$\pm$0.0037} & 0.6622\scriptsize{$\pm$0.0000}\tabularnewline
\textbf{DRAGON} & 0.9101\scriptsize{$\pm$0.0023} & 0.9524\scriptsize{$\pm$0.0020} & 0.8872\scriptsize{$\pm$0.0019} & 0.9418\scriptsize{$\pm$0.0066} & 0.9186\scriptsize{$\pm$0.0063} & 0.9494\scriptsize{$\pm$0.0056} & 0.6124\scriptsize{$\pm$0.0060} & 0.7013\scriptsize{$\pm$0.0060}\tabularnewline
\textbf{DRAGON+} & 0.9201\scriptsize{$\pm$0.0020} & 0.9601\scriptsize{$\pm$0.0037} & 0.8901\scriptsize{$\pm$0.0003} & 0.9539\scriptsize{$\pm$0.0056} & 0.9238\scriptsize{$\pm$0.0061} & 0.9786\scriptsize{$\pm$0.0070} & 0.6200\scriptsize{$\pm$0.0016} & 0.7221\scriptsize{$\pm$0.0025}\tabularnewline
\midrule 
\multicolumn{9}{c}{FTSAD-25}\tabularnewline
\midrule 
PBFM & 0.6712\scriptsize{$\pm$0.0060} & 0.9173\scriptsize{$\pm$0.0008} & 0.7235\scriptsize{$\pm$0.0054} & 0.5571\scriptsize{$\pm$0.0022} & 0.8090\scriptsize{$\pm$0.0048} & 0.7546\scriptsize{$\pm$0.0057} & 0.6110\scriptsize{$\pm$0.0054} & 0.7701\scriptsize{$\pm$0.0048}\tabularnewline
IoTEF & 0.9513\scriptsize{$\pm$0.0012} & 0.9413\scriptsize{$\pm$0.0009} & 0.9828\scriptsize{$\pm$0.0026} & 0.8712\scriptsize{$\pm$0.0040} & 0.9616\scriptsize{$\pm$0.0008} & 0.8654\scriptsize{$\pm$0.0040} & 0.6918\scriptsize{$\pm$0.0039} & 0.8371\scriptsize{$\pm$0.0032}\tabularnewline
TBAFT & 0.9351\scriptsize{$\pm$0.0061} & 0.9547\scriptsize{$\pm$0.0046} & 0.9561\scriptsize{$\pm$0.0064} & 0.8517\scriptsize{$\pm$0.0033} & 0.9554\scriptsize{$\pm$0.0058} & 0.8952\scriptsize{$\pm$0.0041} & 0.6713\scriptsize{$\pm$0.0004} & 0.8127\scriptsize{$\pm$0.0070}\tabularnewline
AWGG & 0.9013\scriptsize{$\pm$0.0089} & 0.9326\scriptsize{$\pm$0.0092} & 0.8913\scriptsize{$\pm$0.0022} & 0.8091\scriptsize{$\pm$0.0079} & 0.9114\scriptsize{$\pm$0.0013} & 0.5104\scriptsize{$\pm$0.0017} & 0.6542\scriptsize{$\pm$0.0026} & 0.7817\scriptsize{$\pm$0.0062}\tabularnewline
TopoMAD & 0.9123\scriptsize{$\pm$0.0019} & 0.9497\scriptsize{$\pm$0.0075} & 0.9781\scriptsize{$\pm$0.0002} & 0.9882\scriptsize{$\pm$0.0065} & 0.9637\scriptsize{$\pm$0.0050} & 0.2622\scriptsize{$\pm$0.0009} & 0.7474\scriptsize{$\pm$0.0011} & 0.8720\scriptsize{$\pm$0.0057}\tabularnewline
StepGAN & 0.9233\scriptsize{$\pm$0.0022} & 0.9378\scriptsize{$\pm$0.0056} & 0.9669\scriptsize{$\pm$0.0028} & 0.9866\scriptsize{$\pm$0.0022} & 0.9521\scriptsize{$\pm$0.0018} & 0.3839\scriptsize{$\pm$0.0023} & 0.7512\scriptsize{$\pm$0.0041} & 0.8654\scriptsize{$\pm$0.0010}\tabularnewline
\textbf{DRAGON} & 0.9781\scriptsize{$\pm$0.0040} & 0.9665\scriptsize{$\pm$0.0075} & 0.9978\scriptsize{$\pm$0.0082} & 0.9918\scriptsize{$\pm$0.0048} & 0.9819\scriptsize{$\pm$0.0008} & 1.0148\scriptsize{$\pm$0.0014} & 0.7981\scriptsize{$\pm$0.0034} & 0.9017\scriptsize{$\pm$0.0053}\tabularnewline
\textbf{DRAGON+} & 0.9812\scriptsize{$\pm$0.0058} & 0.9665\scriptsize{$\pm$0.0010} & 0.9998\scriptsize{$\pm$0.0051} & 0.9934\scriptsize{$\pm$0.0005} & 0.9829\scriptsize{$\pm$0.0007} & 1.0412\scriptsize{$\pm$0.0062} & 0.8013\scriptsize{$\pm$0.0055} & 0.9172\scriptsize{$\pm$0.0003}\tabularnewline
\midrule 
\multicolumn{9}{c}{FTSAD-55}\tabularnewline
\midrule 
PBFM & 0.8912\scriptsize{$\pm$0.0066} & 0.8721\scriptsize{$\pm$0.0048} & 0.9163\scriptsize{$\pm$0.0020} & 0.6029\scriptsize{$\pm$0.0006} & 0.8937\scriptsize{$\pm$0.0061} & 0.8336\scriptsize{$\pm$0.0053} & 0.7612\scriptsize{$\pm$0.0052} & 0.7410\scriptsize{$\pm$0.0051}\tabularnewline
IoTEF & 0.8888\scriptsize{$\pm$0.0011} & 0.9038\scriptsize{$\pm$0.0068} & 0.9424\scriptsize{$\pm$0.0060} & 0.9011\scriptsize{$\pm$0.0078} & 0.9227\scriptsize{$\pm$0.0044} & 0.8304\scriptsize{$\pm$0.0076} & 0.7812\scriptsize{$\pm$0.0025} & 0.7717\scriptsize{$\pm$0.0032}\tabularnewline
TBAFT & 0.7812\scriptsize{$\pm$0.0034} & 0.7673\scriptsize{$\pm$0.0062} & 0.9801\scriptsize{$\pm$0.0065} & 0.9635\scriptsize{$\pm$0.0033} & 0.8607\scriptsize{$\pm$0.0035} & 0.8065\scriptsize{$\pm$0.0080} & 0.7911\scriptsize{$\pm$0.0064} & 0.6526\scriptsize{$\pm$0.0063}\tabularnewline
AWGG & 0.8964\scriptsize{$\pm$0.0029} & 0.9054\scriptsize{$\pm$0.0076} & 0.9121\scriptsize{$\pm$0.0069} & 0.9251\scriptsize{$\pm$0.0019} & 0.9087\scriptsize{$\pm$0.0046} & 0.5089\scriptsize{$\pm$0.0046} & 0.8013\scriptsize{$\pm$0.0026} & 0.7915\scriptsize{$\pm$0.0070}\tabularnewline
TopoMAD & 0.9350\scriptsize{$\pm$0.0024} & 0.9105\scriptsize{$\pm$0.0001} & 0.9212\scriptsize{$\pm$0.0009} & 0.9355\scriptsize{$\pm$0.0001} & 0.9158\scriptsize{$\pm$0.0031} & 0.2491\scriptsize{$\pm$0.0002} & 0.8129\scriptsize{$\pm$0.0051} & 0.7998\scriptsize{$\pm$0.0001}\tabularnewline
StepGAN & 0.9361\scriptsize{$\pm$0.0010} & 0.9128\scriptsize{$\pm$0.0017} & 0.9107\scriptsize{$\pm$0.0031} & 0.9078\scriptsize{$\pm$0.0007} & 0.9117\scriptsize{$\pm$0.0085} & 0.3676\scriptsize{$\pm$0.0011} & 0.8022\scriptsize{$\pm$0.0052} & 0.8041\scriptsize{$\pm$0.0060}\tabularnewline
\textbf{DRAGON} & 0.9440\scriptsize{$\pm$0.0019} & 0.9038\scriptsize{$\pm$0.0005} & 0.9469\scriptsize{$\pm$0.0047} & 0.9916\scriptsize{$\pm$0.0048} & 0.9248\scriptsize{$\pm$0.0046} & 0.9558\scriptsize{$\pm$0.0039} & 0.8452\scriptsize{$\pm$0.0082} & 0.8612\scriptsize{$\pm$0.0042}\tabularnewline
\textbf{DRAGON+} & 0.9561\scriptsize{$\pm$0.0094} & 0.9120\scriptsize{$\pm$0.0068} & 0.9469\scriptsize{$\pm$0.0081} & 0.9927\scriptsize{$\pm$0.0077} & 0.9291\scriptsize{$\pm$0.0080} & 0.9842\scriptsize{$\pm$0.0012} & 0.8671\scriptsize{$\pm$0.0040} & 0.9013\scriptsize{$\pm$0.0079}\tabularnewline
\bottomrule 
\end{tabular}}
\end{table*}

\begin{figure*}[!t]
    \centering
    \includegraphics[width=0.9\linewidth]{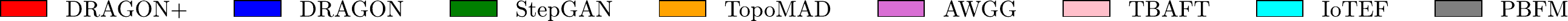} \\
    \subfigure[Energy Consumption]{
    \includegraphics[height=.185\textwidth]{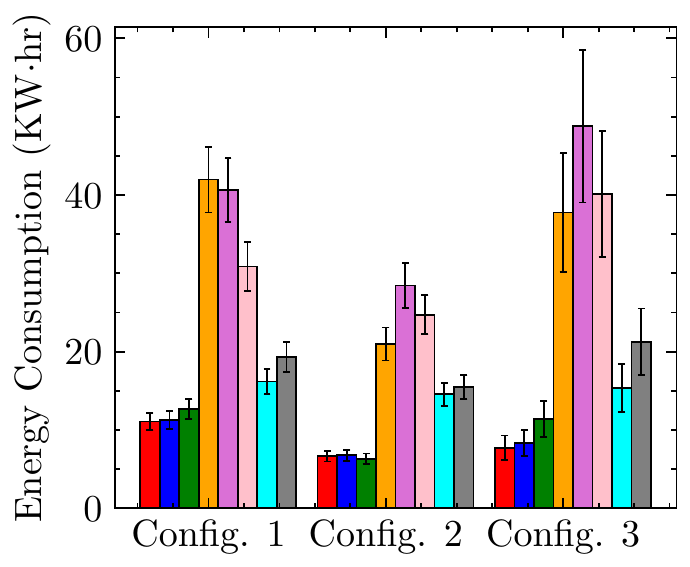}
    \label{fig:energy}
    }
    \subfigure[Response Time]{
    \includegraphics[height=.18\textwidth]{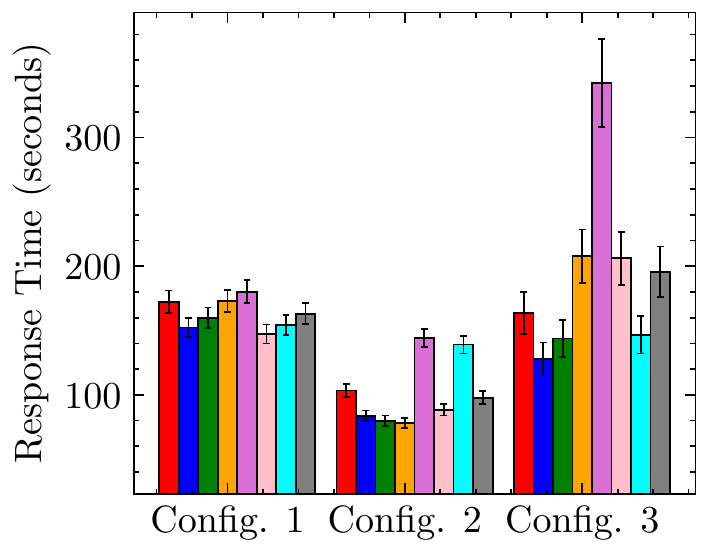}
    \label{fig:response}
    }
    \subfigure[SLO Violation Rate]{
    \includegraphics[height=.18\textwidth]{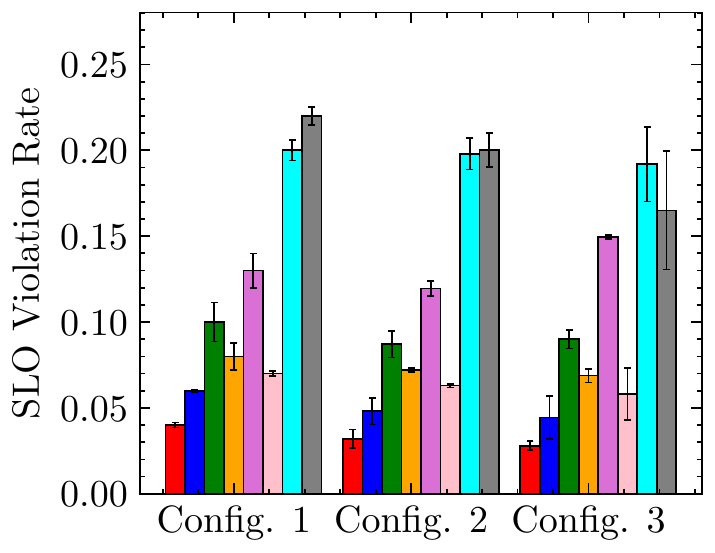}
    \label{fig:sla}
    }
    \subfigure[Fairness]{
    \includegraphics[height=.18\textwidth]{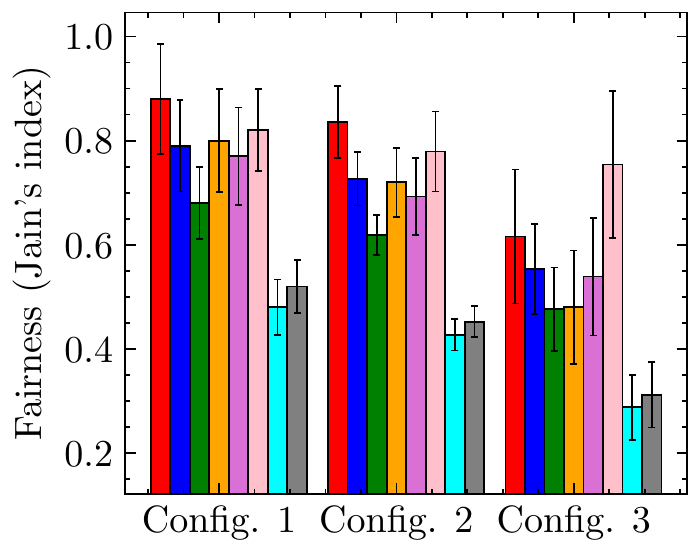}
    \label{fig:fairness}
    }\\ 
    \subfigure[Migration Count]{
    \includegraphics[height=.18\textwidth]{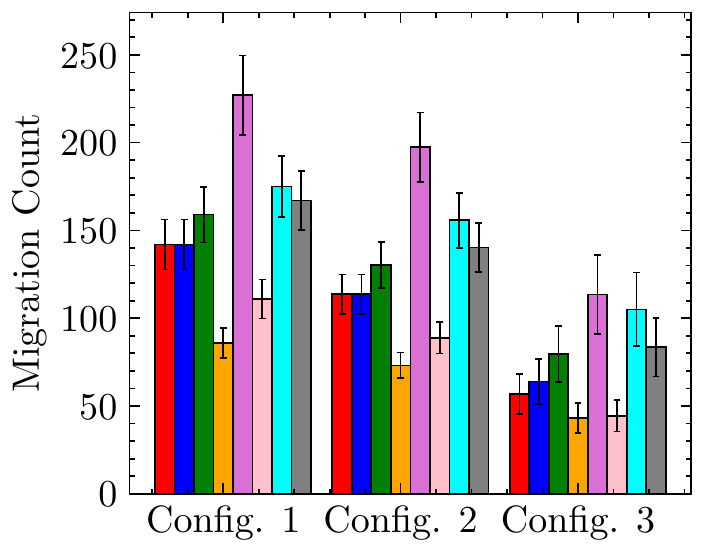}
    \label{fig:mc}
    }
    \subfigure[Migration Time]{
    \includegraphics[height=.18\textwidth]{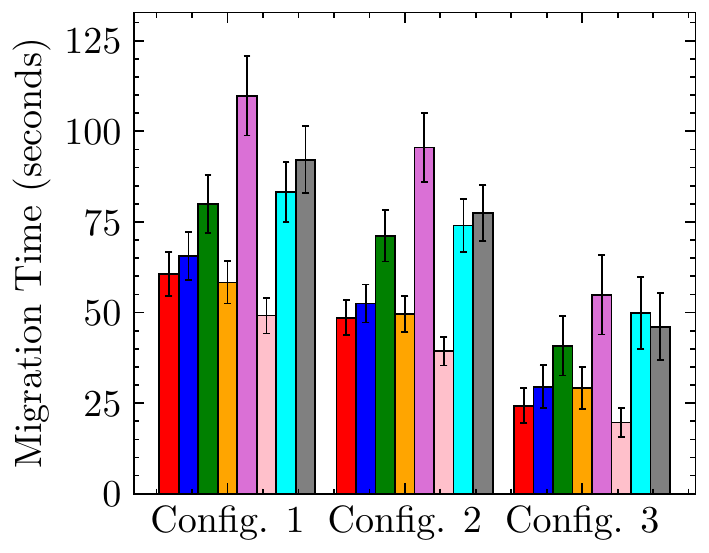}
    \label{fig:mt}
    }
    \subfigure[Scheduling Time]{
    \includegraphics[height=.18\textwidth]{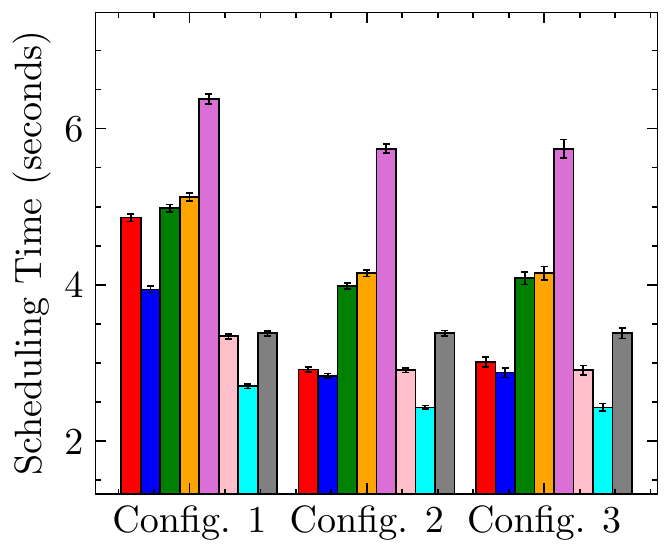}
    \label{fig:scheduling}
    }
    \subfigure[Memory Consumption]{
    \includegraphics[height=.18\textwidth]{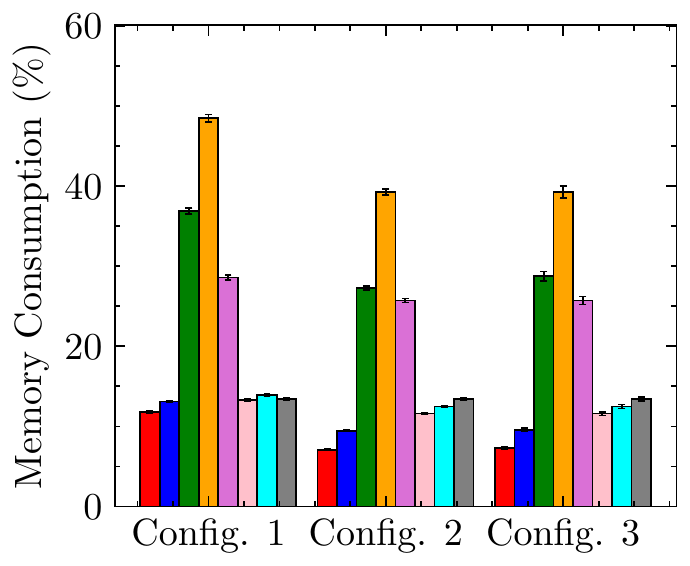}
    \label{fig:memory}
    }
    \caption{Comparison of QoS parameters of DRAGON against baselines. DRAGON and DRAGON+ trained on complete FTSAD-55 data and tested on AIoTBench based workloads.}
    \label{fig:results}
\end{figure*}

\begin{figure*}[!t]
    \centering 
    \includegraphics[width=0.9\linewidth]{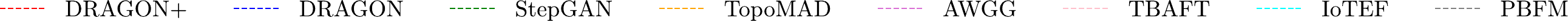}\\
    \includegraphics[width=0.95\linewidth]{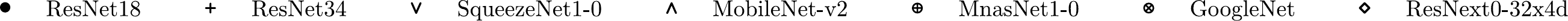}\\
    \subfigure[Response Time]{
    \includegraphics[width=.43\textwidth]{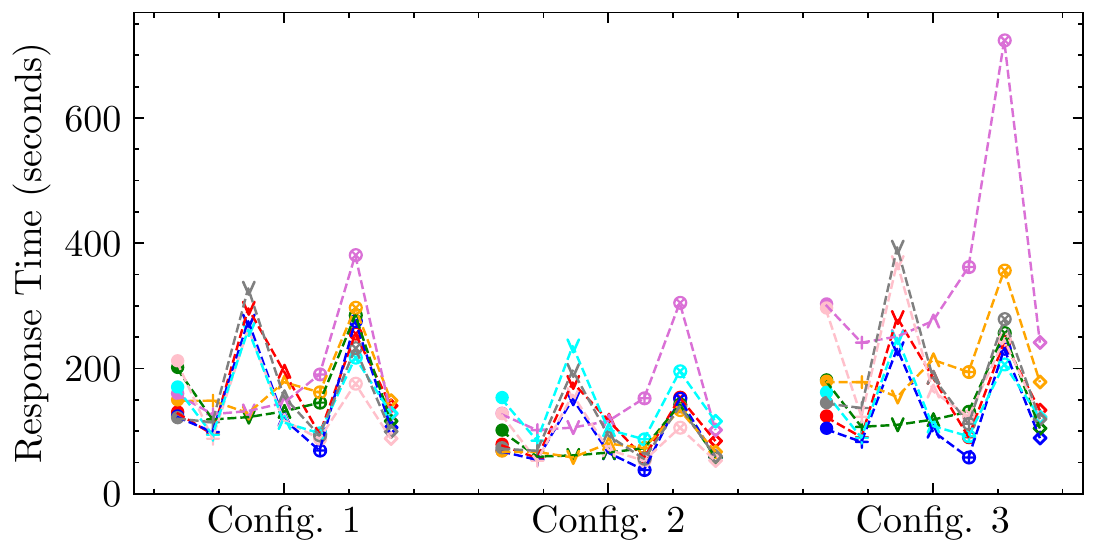}
    \label{fig:response_pa}
    } \hfill
    \subfigure[SLO Violation Rate]{
    \includegraphics[width=.4\textwidth]{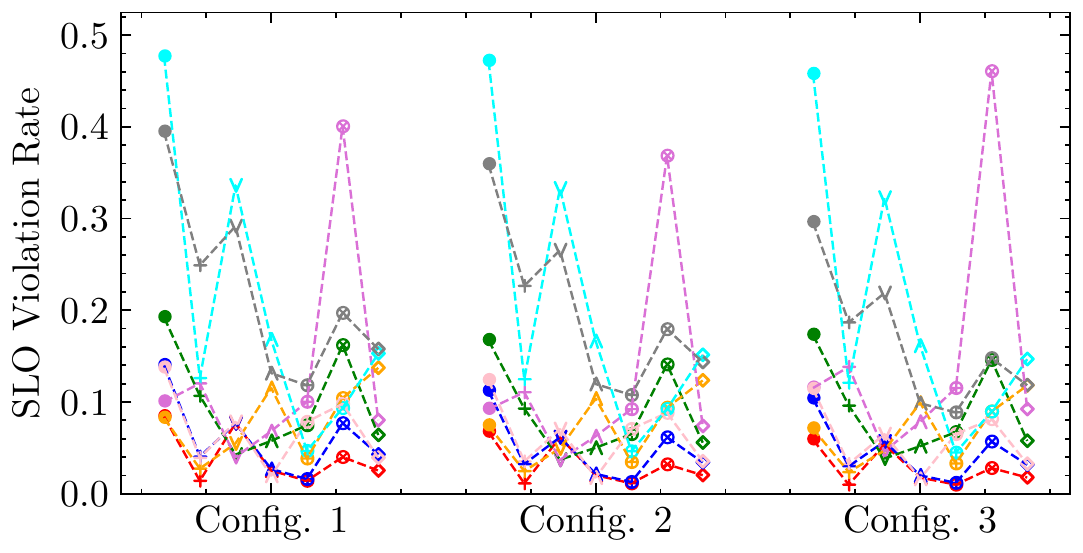}
    \label{fig:sla_pa}
    }
    \caption{Comparison of QoS parameters of DRAGON against baselines for each application type in AIoTBench workloads.} 
    \label{fig:results_pa}
\end{figure*}

\subsection{Results}

\textbf{Comparison of Fault-Detection Performance.} The detection and diagnosis scores for the FTSAD datasets are presented in Table~\ref{tab:detection}. We do not present results on AIoTBench as it does not have fault labels. We use the \textit{point-adjust} technique to determine whether a fault prediction is true or false. Specifically, faulty observations usually occur in the form of contiguous anomaly segments; thus, if at least one observation of an faulty time-series window is correctly detected, all the other observations of the segment are also considered as correct. The approach was first presented in~\cite{usad}. Compared to the baselines, the DRAGON and DRAGON+ models improve upon the accuracy, precision and recall by up to 17.28\%-11.87\%-14.14\% and 18.32\%-12.50\%-14.34\% respectively. Among the baselines, the StepGAN baseline has the highest average F1 score of 0.9040. The DRAGON and DRAGON+ models gives a higher average F1 scores of 0.9418 and 0.9453, \textit{i.e.}, 4.18\% and 4.56\% higher than StepGAN respectively. In terms of AUROC, the TopoMAD baseline has the highest value of 0.9416. The proposed models outperform this score by 3.56\%-4.08\%, getting 0.9751 and 0.9800 scores. When comparing F1/GB, the deep learning methods have the lowest scores. Among the baselines, the TBAFT model has the highest average F1/GB score of 0.8252. DRAGON and DRAGON+ improve upon this by 17.94\%-21.34\%. Compared to the deep learning baselines, DRAGON and DRAGON+ improve the F1/GB by up to 305.03\% and 316.68\% respectively. In terms of diagnosis scores, StepGAN is the best baseline with HR of 0.7160 and NDCG of 0.7772. Compared to these scores, DRAGON and DRAGON give 5.01\%-6.54\% and 5.69\%-8.95\% improvement for HR and NDCG respectively. These results show that having only a single discriminator network allows us to significantly reduce the network's number of parameters and give a performance boost compared to having two such networks in StepGAN.

\begin{figure*}[t]
    \centering 
    \includegraphics[width=.85\textwidth]{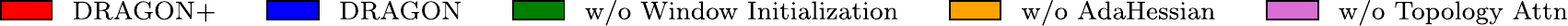}\\
    \subfigure[F1 Score]{
    \includegraphics[height=.170\textwidth]{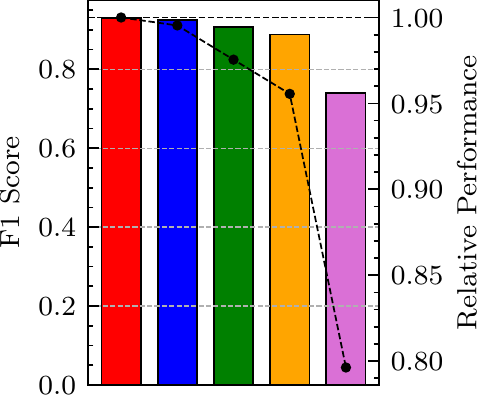}
    \label{fig:a_f1}
    }
    \subfigure[Train Time]{
    \includegraphics[height=.170\textwidth]{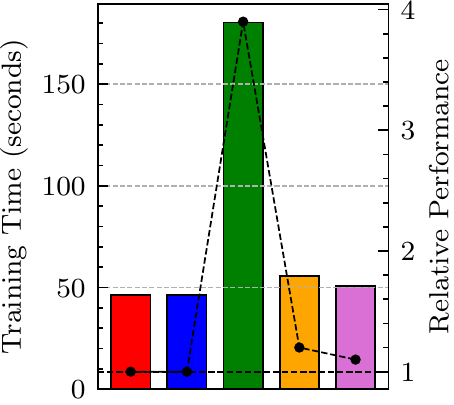}
    \label{fig:a_traintime}
    }
    \subfigure[Memory]{
    \includegraphics[height=.170\textwidth]{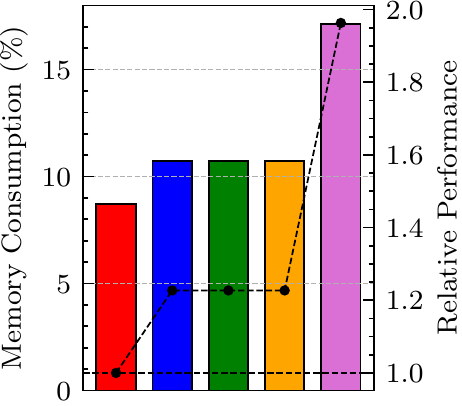}
    \label{fig:a_memory}
    }
    \subfigure[Scheduling Time]{
    \includegraphics[height=.170\textwidth]{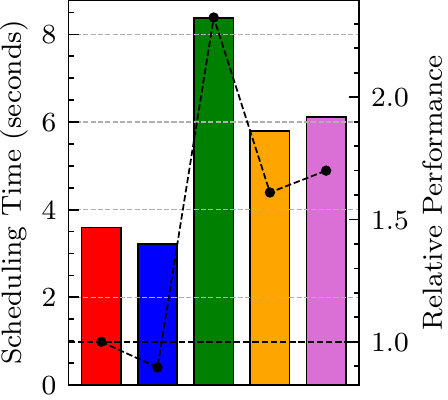}
    \label{fig:a_sched}
    } 
    \subfigure[Energy]{
    \includegraphics[height=.170\textwidth]{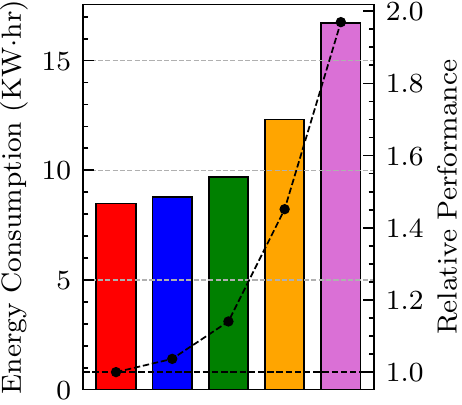}
    \label{fig:a_energy}
    }
    \subfigure[Response Time]{
    \includegraphics[height=.170\textwidth]{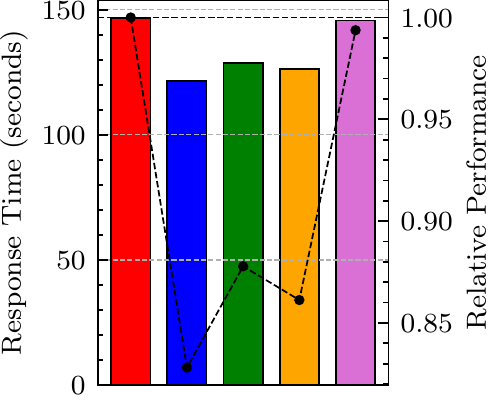}
    \label{fig:a_response}
    }
    \subfigure[SLO Violations]{
    \includegraphics[height=.170\textwidth]{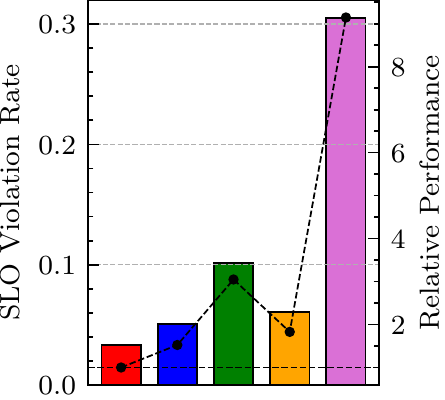}
    \label{fig:a_sla}
    }
    \caption{Ablation Analysis of different model components of DRAGON. The bar graphs show absolute values. The line graphs show the performance relative to DRAGON+. F1 scores and training times are reported on FTSAD-55 dataset, while the QoS results are reported on the AIoTBench tests averaged across the three system configurations.}
    \label{fig:ablation}
\end{figure*}

\begin{figure*}[t]
    \centering 
    \includegraphics[width=.3\textwidth]{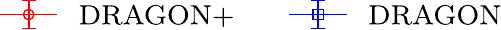}\\
    \subfigure[F1 Score]{
    \includegraphics[height=.170\textwidth]{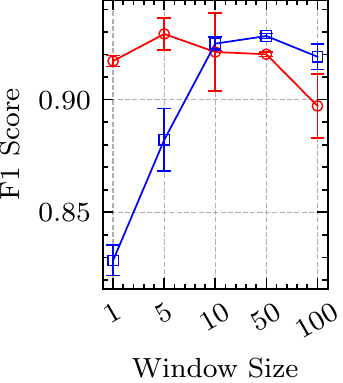}
    \label{fig:w_f1}
    }
    \subfigure[Train Time]{
    \includegraphics[height=.170\textwidth]{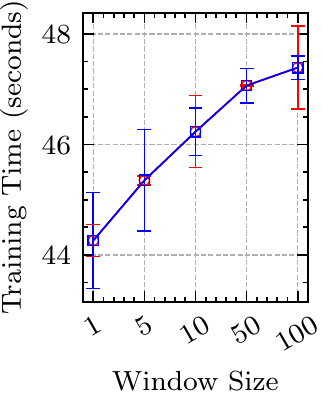}
    \label{fig:w_traintime}
    }
    \subfigure[Memory]{
    \includegraphics[height=.170\textwidth]{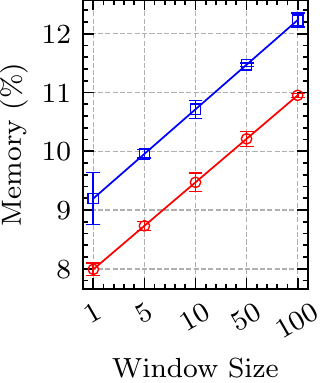}
    \label{fig:w_memory}
    } 
    \subfigure[Scheduling Time]{
    \includegraphics[height=.170\textwidth]{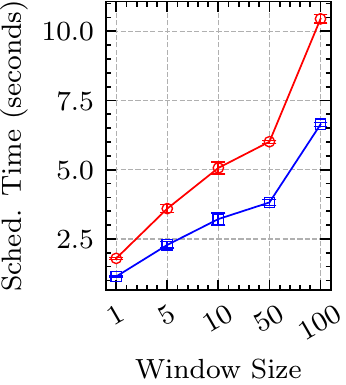}
    \label{fig:w_sched}
    } 
    \subfigure[Energy]{
    \includegraphics[height=.170\textwidth]{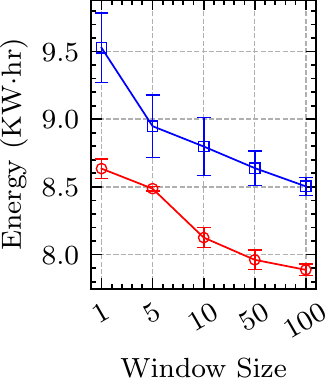}
    \label{fig:w_energy}
    }
    \subfigure[SLO Violations]{
    \includegraphics[height=.170\textwidth]{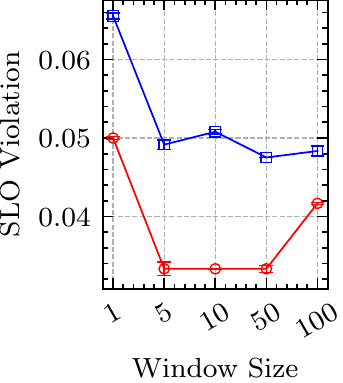}
    \label{fig:w_sla}
    }
    \caption{Sensitivity Analysis of performance of DRAGON and DRAGON+ with window size. DRAGON+ has highest F1 score for window size of 5, whereas DRAGON has it for window size of 10.  F1 scores and training times are reported on FTSAD-55 dataset, while the QoS results are reported on the AIoTBench tests averaged across the three system configurations.}
    \label{fig:sens_w}
\end{figure*}

\begin{figure*}[t]
    \centering 
    \includegraphics[width=.3\textwidth]{sens/legend.pdf}\\
    \subfigure[F1 Score]{
    \includegraphics[height=.168\textwidth]{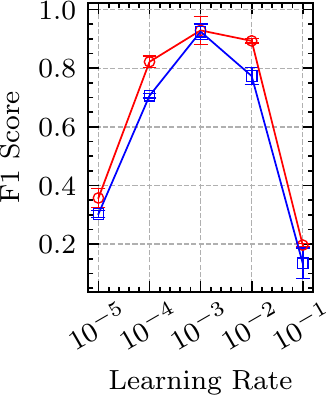}
    \label{fig:lr_f1}
    }
    \subfigure[Train Time]{
    \includegraphics[height=.168\textwidth]{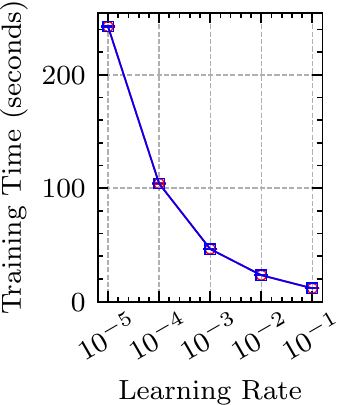}
    \label{fig:lr_traintime}
    }
    \subfigure[Memory]{
    \includegraphics[height=.168\textwidth]{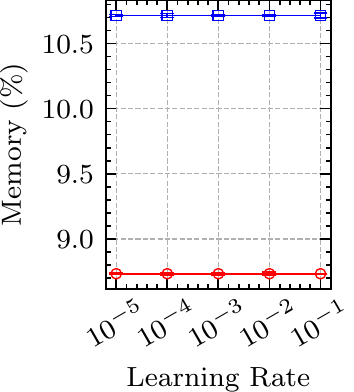}
    \label{fig:lr_memory}
    } 
    \subfigure[Scheduling Time]{
    \includegraphics[height=.168\textwidth]{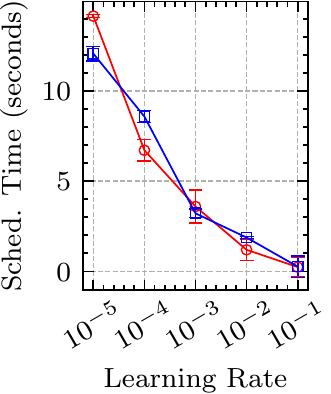}
    \label{fig:lr_sched}
    } 
    \subfigure[Energy]{
    \includegraphics[height=.168\textwidth]{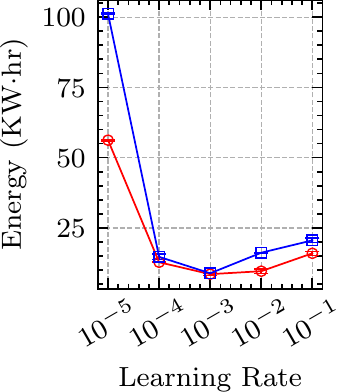}
    \label{fig:lr_energy}
    }
    \subfigure[SLO Violations]{
    \includegraphics[height=.168\textwidth]{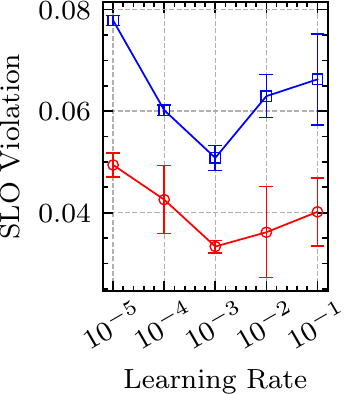}
    \label{fig:lr_sla}
    }
    \caption{Sensitivity Analysis of performance of DRAGON and DRAGON+ with AdaHessian learning rate. Both have highest scores for learning rate of $10^{-3}$.  F1 scores and training times are reported on FTSAD-55 dataset, while the QoS results are reported on the AIoTBench tests averaged across the three system configurations.}
    \label{fig:sens_lr}
\end{figure*}

\textbf{Comparison of QoS Metrics.} We now present the QoS scores of all models for the three federated configurations in  Fig.~\ref{fig:results}. Fig.~\ref{fig:results_pa} presents the average response time and SLO violation rates for each application in the AIoTBench workloads. To test DRAGON and DRAGON+, we use a GON network trained on the FTSAD-55 dataset and the scheduling decisions from the state-of-the-art GOBI scheduler~\cite{tuli2021cosco}. Training time of the GON network on this dataset is 46.23 seconds. Among the baselines, StepGAN has the lowest energy consumption of 10.15 KW-hr averaged over the three configurations. The DRAGON+ model has the lowest energy consumption of 8.48 KW-hr, 16.40\% lower than StepGAN. Second lowest is DRAGON with 8.79 KW-hr, 13.39\% lower than StepGAN (Fig.~\ref{fig:energy}). We see similar trends for response time. The AWGG model struggles in the third configuration with heterogeneous resource capacities across LEIs due to the drop in the performance of reconstructive clustering. DRAGON and DRAGON+ have average response times comparable to the best baseline for each configuration.  Comparing the energy consumption and response times across configurations, we observe that they are lower for the second configuration compared to the first. This is primarily because the second configuration has only 4 running instances of the fault-tolerance model due to four brokers. The third configuration has the highest energy consumption and response time due to the increasing difficulty in fault-tolerance in heterogeneous federated setups (Figs.~\ref{fig:energy} and~\ref{fig:response}). The SLO violation rates remain mostly consistent across configurations, except for AWGG that has a higher violation rate for the third configuration due to higher response times (Figs.~\ref{fig:sla} and~\ref{fig:sla_pa}). Among all baselines, TBAFT has the lowest average SLO violation rate of 6.37\%, where DRAGON and DRAGON+ improve upon this by 20.25\% and 47.72\%, giving violation rates of 5.08\% and 3.33\% respectively (Fig.~\ref{fig:sla}). In terms of fairness, DRAGON+ has the highest average fairness index of 0.78 (Fig.~\ref{fig:fairness}). The fairness drops from the first to third configurations due to more disparate LEI resources. In terms of migration count and migration time, AWGG is the most aggressive in terms of migrations~(Figs.~\ref{fig:mc} and~\ref{fig:mt}). On the other hand, TopoMAD is the most parsimonious. DRAGON and DRAGON+ trade off between the fault-tolerance of migrations and their overheads. In terms of scheduling time, the heuristic solutions have the lowest times with IoTEF having only 2.52 seconds on an average. The deep learning methods have much higher scheduling times. However, for the second and third configurations, deep learning baselines have scheduling times consistent with the first configuration. On the other hand, DRAGON and DRAGON+ have lower scheduling times than all deep learning baselines. This is due to the dot-product attention in the GON network and the higher-order optimization, allowing the model to scale with the dimension size of the time-series window and the number of edge devices in an LEI (Fig.~\ref{fig:scheduling}). Memory consumption of the DRAGON and DRAGON+ models is the lowest, being only 10.71\% and 8.72\%. It is lower for DRAGON+ due to the lower parameter size GON network compared to DRAGON (see Section~\ref{sec:impl} and Fig.~\ref{fig:memory}).

\blue{The DRAGON+ method is not meant to substitute DRAGON, but rather offer a different tradeoff of latency and detection accuracy. Comparing DRAGON+ and DRAGON, we observe that the former has a higher average response time compared to the latter. This is due to the multi-step lookeahead and optimization of the migration decision. However, the more informed decision choice in DRAGON+ leads to improvements in other metrics such as energy consumption, SLO violation rates and memory consumption. Although both methods beat state-of-the-art baselines, the above demonstrates that DRAGON is more suitable for latency critical decision making or when the broker nodes are resource constrained. However, for resource abundant brokers or when energy or SLO violation rates have higher importance, DRAGON+ is more appropriate.}

\subsection{Ablation Study}

To study the relative importance of each component of the model, we exclude every major component one at a time and observe how it affects the performance of the scheduler. An overview of this ablation analysis is given in Fig~\ref{fig:ablation}. First, we consider the DRAGON+ and DRAGON models. We then consider DRAGON with random noise initialization of $Z$ instead of $W$ (\textit{w/o Window Initialization} model). Next, we consider a model with the \texttt{Adam} optimizer instead of \texttt{AdaHessian} (\textit{w/o AdaHessian} model). Lastly, we consider the DRAGON model without the topology attention, using only $W$ and $S$ as inputs (\textit{w/o Topology Attn} model). We report the following findings:

\begin{itemize}[leftmargin=*]
    \item Without the multi-step calculation of the fitness score in DRAGON+, the DRAGON model performs a myopic optimization and suffers an increase in the  SLO violation rate, energy consumption and memory footprint.
    \item Without the window initialization, the generation stage significantly more time, increasing the scheduling and training time of the model.
    \item Without second-order optimization, the training and scheduling times are much higher.
    \item Without the topology attention, the SLO violation rate drastically increases as a model in a broker does not have any contextual information of the rest of the federated setup.
\end{itemize}
We observe that window initialization and topology attention have the maximum impact to the performance. 

\subsection{Sensitivity Analysis}
\label{sec:sensitivity}

Figs.~\ref{fig:sens_w} and~\ref{fig:sens_lr} provide a sensitivity analysis of the performance of DRAGON and DRAGON+ models with window size and \texttt{AdaHessian} learning rate respectively. This sensitivity analysis is done to highlight the trade-off between the model performance and training time as we increase either of these parameters. As window size increases, the memory consumption, training time and scheduling time jointly increase due to larger input size of the GON network giving rise to higher time to generate next time-series window. However, higher window size also allows the model to get a broader context of the environment, reducing energy consumption and SLO violation rates. For DRAGON, the F1 score increases monotonically until a window size of 10, after which it remains mostly consistent. However, for DRAGON+, increasing the window size deteriorates the F1 score after 5. Intuitively, this is because DRAGON+ is able to get a broader temporal context due to multi-step iterations, and larger windows make model training unstable. Thus, we choose window size of 10 for DRAGON and 5 for DRAGON+ in our experiments (mentioned in Section~\ref{sec:impl}).

The variation of performance metrics with the \texttt{AdaHessian} learning rate is more straightforward. Memory consumption is independent of the learning rate of the optimizer. However, training time and scheduling time of DRAGON and DRAGON+ decrease as we increase the learning rate. This is due to larger jumps in the optimization steps. However, for a high learning rate of $\geq 10^{-2}$, the model is unable to converge to the optima and hence we see the increase in energy consumption and SLO violation rates for these values. In terms of F1 score, it is highest for learning rate of $10^{-3}$ and hence has been used in our experiments.

\section{Conclusions}
\label{sec:conclusion}
We have presented a decentralized fault-tolerance model, DRAGON, that utilizes fault predictions from GON and runs co-simulated fictitious play to converge to preemptive migration decisions. DRAGON leverages a novel memory-efficient generative model called generative optimization networks (GONs). Unlike previously proposed GAN models, GON uses a single neural network to both discriminate and generate samples. This comes with the advantage of significant savings in terms of memory footprint. Unlike prior work aiming at reducing the parameter size of popular generative models such as GANs, this framework is suitable for memory-constrained systems like edge devices. We extend DRAGON to DRAGON+ where we perform a multi-step optimization of a fitness score. The proposed models give up to 4.56\% higher F1 scores compared to the best baseline StepGAN. In a federated Raspberry Pi based edge setup with real-world AIoTBench workloads, DRAGON and DRAGON+ are able to outperform baselines by giving up to 74\%, 63\% and 82\% better energy consumption, response time and service deadline violations, respectively. These improvement are primarily due to advances like window initialization, second-order optimization and graph topology based attention.

In the future, the model could be adapted to make it more robust against non-altruistic entities in the system. \blue{Further, the preemptive migration decisions could be taken using a deep-reinforcement learning approach, such as deep deterministic policy gradient, which is informed by a fault-detection mechanism. The detection and decision making could be combined for holistic QoS optimization and further reduction in memory consumption or decision latency.}

\section*{Software Availability}
The code and relevant training scripts are available on GitHub under BSD-3 licence at \url{https://github.com/imperial-qore/DRAGON}.

\section*{Acknowledgments}
Shreshth Tuli is supported by the President’s Ph.D. Scholarship at the Imperial College London.





\bibliographystyle{IEEEtran}
\bibliography{references}

\end{document}